\documentclass[preprint,12pt,authoryear]{elsarticle}

\usepackage{amssymb}
\newcommand{\eg}{{\em e.g.}}           % e.g.
\newcommand{\ie}{{\em i.e.}}           % i.e.
\usepackage{mathrsfs}
\usepackage[normalem]{ulem}
\usepackage[symbol]{footmisc}
\usepackage{subfig}
\usepackage{bbm}
\usepackage{colortbl}
\definecolor{mygray}{gray}{0.9}

\usepackage{cite}
\usepackage{booktabs,tabularx}
\usepackage{array, multirow}
\usepackage{colortbl}
\definecolor{mygray}{gray}{0.9}

\usepackage{pifont}% http://ctan.org/pkg/pifont
\newcommand{\x}{\mathbf{X}}
\newcommand{\y}{\mathbf{Y}}
\newcommand{\s}{\mathcal{S}}

\usepackage{tikz}
\newcommand*\emptycirc[1][1ex]{\tikz\draw (0,0) circle (#1);} 
\newcommand*\halfcirc[1][1ex]{%
  \begin{tikzpicture}
  \draw[fill] (0,0)-- (90:#1) arc (90:270:#1) -- cycle ;
  \draw (0,0) circle (#1);
  \end{tikzpicture}}
\newcommand*\fullcirc[1][1ex]{\tikz\fill (0,0) circle (#1);} 

\usepackage{algorithm,algorithmic}
\usepackage{multirow}
\usepackage{rotating}
\usepackage{amsmath,amsfonts,bm}
\usepackage{hyperref}

\DeclareMathOperator*{\argmin}{\arg\min}

\journal{Medical Image Analysis}

\begin{document}

\begin{frontmatter}



\title{FIESTA: Fourier-Based Semantic Augmentation with Uncertainty Guidance for Enhanced Domain Generalizability in Medical Image Segmentation}

\author[inst1]{Kwanseok Oh}
\author[inst2]{Eunjin Jeon}
\author[inst1]{Da-Woon Heo}
\author[inst1]{Yooseung Shin}
\author[inst1,inst2]{Heung-Il Suk\textsuperscript{$\dagger$,}}
\affiliation[inst1]{organization={Department of Artificial Intelligence, Korea University},addressline={Seoul 02841, Republic of Korea}}
\affiliation[inst2]{organization={Department of Brain and Cognitive Engineering, Korea University},addressline={Seoul 02841, Republic of Korea}}

% Sentences should avoid starting with informal terms at the beginning in academic writing. Avoid using also, besides, so, and, meanwhile, or, and but at the beginning. Consider using however, further, furthermore, moreover, in addition, additionally, nonetheless, or nevertheless.

\begin{abstract}
\footnotetext[2]{Corresponding author: Heung-Il Suk (hisuk@korea.ac.kr)}
Single-source domain generalization (SDG) in medical image segmentation (MIS) aims to generalize a model using data from only one source domain to segment data from an unseen target domain. Despite substantial advances in SDG with data augmentation, existing methods often fail to fully consider the details and uncertain areas prevalent in MIS, leading to mis-segmentation. This paper proposes a Fourier-based semantic augmentation method called FIESTA using uncertainty guidance to enhance the fundamental goals of MIS in an SDG context by manipulating the amplitude and phase components in the frequency domain. The proposed Fourier augmentative transformer addresses semantic amplitude modulation based on meaningful angular points to induce pertinent variations and harnesses the phase spectrum to ensure structural coherence. Moreover, FIESTA employs epistemic uncertainty to fine-tune the augmentation process, improving the ability of the model to adapt to diverse augmented data and concentrate on areas with higher ambiguity. Extensive experiments across three cross-domain scenarios demonstrate that FIESTA surpasses recent state-of-the-art SDG approaches in segmentation performance and significantly contributes to boosting the applicability of the model in medical imaging modalities.
\end{abstract}


% [Don't try to capture all ideas, concepts or conclusions as highlights are meant to be short: 85 characters or fewer, including spaces.]
\begin{highlights}
\item FIESTA enhances single-source domain generalization in medical image segmentation 
\item Modulates amplitude and phase components, advancing model generalizability
\item Incorporates epistemic uncertainty in FIESTA, guiding augmentation refinement
\item Demonstrates state-of-the-art outcomes in three challenging cross-domain scenarios
\item Verifies the scalability of FIESTA by applying it to the segment anything model
% \item Research highlights 1: FIESTA for enhanced single-source domain generalization in medical image segmentation % 85
% \item Research highlights 2: Modulating amplitude and phase components in FAT, advancing model's generalizability % 84
% \item Research highlights 3: Incorporating epistemic uncertainty in FIESTA to guide the refinement of augmentation % 85
% \item Research highlights 4: Demonstrated state-of-the-art outcomes in three challenging cross-domain scenarios % 82
% \item Research highlights 5: Verification of FIESTA's scalability by applying it to the Segment Anything model % 81
\end{highlights}


\end{frontmatter}


% \clearpage
\section{Introduction}\label{sec:intro}
Medical image segmentation (MIS) is a significant, pivotal challenge in medical image analysis, empowering clinicians to delineate specific anatomical regions by isolating the imperative region of interest from the overall image \citep{pham2000current,sharma2010automated}. Advances in deep learning (DL)-based MIS approaches have recently achieved impressive segmentation performance across a spectrum of imaging modalities, including magnetic resonance imaging (MRI), computed tomography (CT), and X-ray \citep{hatamizadeh2022unetr,su2023rethinking,zhou2023nnformer}. However, the success of DL models relies on the assumption that the data in the training and testing phases are drawn from an independent and identically distributed (\ie, i.i.d) set within a singular domain \citep{vapnik1991principles,ben2010theory}. When such an indispensable assumption is violated or unsatisfied even slightly, severe performance degradation occurs, called domain shift \citep{zhou2022domain}. In practice, most medical application scenarios suffer from domain shifts due to the variances between training and testing datasets. These shifts are induced by diverse device manufacturers \citep{liu2021feddg}, imaging protocols \citep{zhuang2022cardiac}, and image modalities \citep{kavur2021chaos}. Such heterogeneous factors cause discrepancies in data distribution, posing a barrier to developing robust models capable of performing consistently across clinical settings.

Previous studies have proposed appealing strategies to enhance the generalizability of DL models to address this problem. Unsupervised domain adaptation (UDA) techniques have emerged to employ labeled data from a source domain with unlabeled data from a target domain to bridge the distribution gap \citep{xie2022unsupervised,shin2023sdc}. Recently, source-free domain adaptation (SFDA) has gained attention due to its unique approach, which is a variant of the UDA \citep{yang2022source,yu2023source,stan2024unsupervised}. Particularly, SFDA-based models are beneficial when source data are restricted owing to privacy or security concerns because they are adapted to the target domain without requiring access to the source domain data during adaptation. Even if UDA and SFDA exhibit promising outcomes on out-of-distribution data, employing these approaches has drawbacks, making it impractical. For seamless adaptation to the target domain distribution, the UDA and SFDA methods demand that data in the target domain be accessible and available during training. Thus, such prerequisites potentially incur difficulties in their deployment in the real world. 

As an alternative approach, multi-source domain generalization does not require unseen target domains because this approach incorporates domain knowledge from multiple source domains during the learning process. Thus, multi-source domain generalization is performed without direct exposure to the target domain, sidestepping the practical problems encountered with SFDA and UDA. However, if the diversity of source domains is scarce or the simulated domain falls short of properly encompassing the distribution of unseen target domains, this approach may compromise performance.

\begin{table}[t!]\centering\scriptsize\setlength{\tabcolsep}{3. pt}
\renewcommand{\arraystretch}{1.2}
    \caption{Primary characteristic summary of each technique.}
    \centering \label{tab:method-category}
    \begin{tabular}{ccccc}
    \toprule
    \multicolumn{1}{c}{\multirow{2}{*}{\textbf{Category}}} & \multirow{1}{*}{\textbf{Unsupervised}} & \multirow{1}{*}{\textbf{Source-Free}} & \multirow{1}{*}{\textbf{Multi-Source}} & \multirow{1}{*}{\textbf{Single-Source}}\\
    & \multirow{1}{*}{\textbf{DA (UDA)}} & \multirow{1}{*}{\textbf{DA (SFDA)}} & \multirow{1}{*}{\textbf{DG (MDG)}} & \multirow{1}{*}{\textbf{DG (SDG)}}\\
    \midrule
    \textbf{Training with source domain(s)} & \fullcirc & \emptycirc & \fullcirc & \fullcirc\\
    % \cmidrule(lr){1-5}
    \textbf{Training with target domain(s)} & \halfcirc & \halfcirc & \emptycirc & \emptycirc\\
    % \cmidrule(lr){1-5}
    \textbf{Enhanced privacy protection} & \emptycirc & \fullcirc & \emptycirc & \halfcirc\\
    % \cmidrule(lr){1-5}
    \textbf{Overfitting risk} & \halfcirc & \halfcirc & \fullcirc & \emptycirc\\
    % \cmidrule(lr){1-5}
    \textbf{Challenge of generalization} & \emptycirc & \emptycirc & \halfcirc & \fullcirc\\
    % \cmidrule(lr){1-5}
    \textbf{Feasibility in practice} & \emptycirc & \halfcirc & \halfcirc & \fullcirc\\
    \bottomrule
    \end{tabular}
    
    \vspace{0.1cm}
    \text{\emptycirc: not available; \halfcirc: partially involved; \fullcirc: completely involved.}\\
    \bigskip
    \begin{tabular}{p{13.5cm}}
    \footnotesize
    Notes: DA: domain adaptation, UDA: unsupervised domain adaptation, SFDA: source-free domain adaptation, DG: domain generalization, MDG: multi-source domain generalization, SDG: single-source domain generalization.
    \end{tabular}
\end{table}

This study delves into a more extreme scenario: single-source domain generalization (SDG) \citep{zhou2022domain}, considering its feasibility for practical applications. Regarding generalization, SDG is suitable when access to target domain instances is either limited or nonexistent because this approach is trained on only one source domain and learns to perform well on multiple unseen target domains (Table \ref{tab:method-category}). Due to insufficient training data, most SDG-based methods have proposed augmentation solutions in the data \citep{devries2017improved,xu2020robust} or feature representation space \citep{huang2020self, zhou2021domain} to address the lack of data diversity. This objective could be encouraged in the field of MIS, where inherent mutability in medical imaging techniques and protocols can drastically influence the model's performance \citep{yoon2023domain}, and obtaining the annotated data is time-consuming and expensive. 

In this context, recent works \citep{zhou2022generalizable,ouyang2022causality,su2023rethinking} have gradually explored innovative augmentation techniques tailored specifically to medical images. While their efforts have contributed to enhancing domain generalizability, they have failed to reflect the intricacies of MIS during augmentation. 
Despite the fact that precise identification of organ delineation or tissue boundaries is paramount in alleviating mis-segmentation in MIS \citep{lee2020structure,liu2021feddg}, research has predominantly concentrated on confined augmentation of the global texture \citep{zhou2022generalizable,ouyang2022causality} or class-specific areas of an image within the ground truth \citep{su2023rethinking}. These approaches further apply saliency-based extra augmentation, and their effectiveness is determined by the linearly interpolated \citep{ouyang2022causality} or model gradient-based \citep{su2023rethinking} saliency. In these cases, model attention is focused on regions completely unrelated to segmentation or with the most significant influence on model performance, potentially overlooking other critical areas. Thus, such lopsided methods may not provide insight to distinguish imperceptible or unclear segment locations, including incorrectly segmented areas, that significantly affect the segmentation quality. In this light, we hypothesize that \textit{the key to successful SDG in MIS lies in enriching the representation diversity, encompassing a broader range of domain variabilities while factoring in semantic appearance and uncertain areas for accurate segmentation.}

\begin{figure}[t]
    \centering
    \includegraphics[width=1\linewidth]{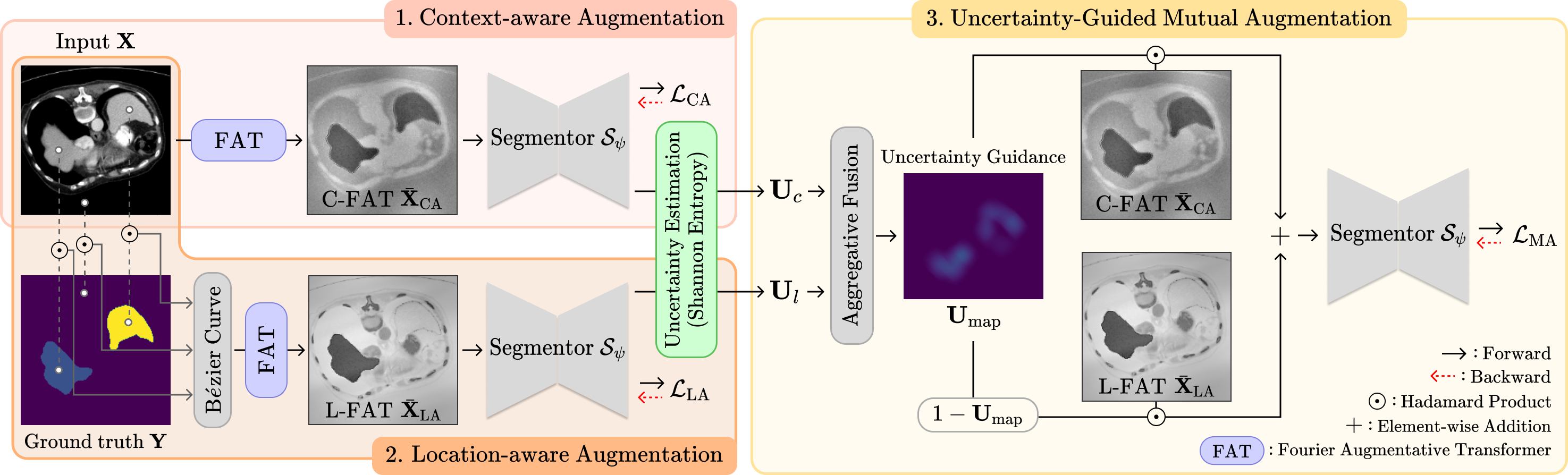}
    \caption{Schematic overview of FIESTA. Context-aware augmentation provides diverse changes throughout the global context via the proposed Fourier augmentative transformer (FAT), whereas location-aware augmentation uses the FAT but additionally adjusts the segment-specific location styles using the B$\Acute{e}$zier curve. Building on these two augmentation phases, uncertainty-guided mutual augmentation further enforces segmentation in ambiguous regions by generating augmented images via uncertainty guidance.}
    \label{fig:FIESTA-framework}
\end{figure}

Based on these propositions, we propose a \textbf{F}our\textbf{IE}r-based \textbf{S}eman\textbf{T}ic \textbf{A}ugmentation method with uncertainty guidance (UG) called FIESTA (Fig. \ref{fig:FIESTA-framework}). First, a simple, efficient data augmentation is designed via the Fourier transform by manipulating the amplitude and phase components in the frequency domain. The concept was motivated by insight from visual psychophysics \citep{piotrowski1982demonstration,guyader2004image,chen2021amplitude}, which suggests that the amplitude spectrum predominantly encodes the \textit{low-level modality awareness of images} (\eg, domain-variant textures), whereas the phase spectrum is capable of capturing \textit{high-level semantic information} (\eg, domain-invariant superficial patterns). 

As a crux module within the FIESTA framework, the Fourier augmentative transformer (FAT) was devised. This FAT imposes two cutting-edge techniques, novel masking and intra-modulation, to transform the amplitude spectrum. For this purpose, the angular density distribution from the series of manipulated amplitudes is calculated in such a way that meaningful directional angles are estimated and used as criteria for mask generation and intra-modulation. The FAT further engages a refined phase attention mechanism that applies a filtering strategy to the phase component. This approach reflects the structural coherence to promote augmentation regarding depicting segments for explicit parcellation (Fig. \ref{fig:fat-framework}). Such a dual focus on the amplitude and phase components via FAT ensures that FIESTA is aligned with the primitive objectives of MIS in SDG, enhancing the robustness and proficiency of the model in portraying precise segmentation under diverse cross-domain scenarios. Moreover, a novel uncertainty-guided mutual augmentation technique is developed by reciprocally exploiting two view images (\ie, context-aware and location-aware augmentation) derived based on the FAT. This strategic incorporation of uncertainty can enable the model to identify and emphasize ambiguous boundaries of anatomies and boost augmentation by amplifying the diversity of these uncertain areas (refer to Fig. \ref{fig:uncer-res}). The primary contributions of the proposed method are summarized as follows:

\begin{itemize}
    \item To the best of our knowledge, this work is the first Fourier-based augmentation method that simultaneously manipulates amplitude and phase components using meaningful factors tailored to SDG for cross-domain MIS.
    \item We propose the FAT, providing an advanced augmentation strategy that combines masking and modulation techniques to transform the amplitude spectrum and applies filtering to refine the phase information to impose structural integrity.
    \item The FIESTA framework embraces an uncertainty-guided mutual augmentation strategy by applying UG to focus learning in the segmentation model on certain areas of high ambiguity or mis-segmented locations.
    \item Based on the quantitative and qualitative experimental results on various cross-domain scenarios (including cross-modality, cross-sequence, and cross-sites), we demonstrate the significant robustness and generalizability of FIESTA, which surpasses state-of-the-art SDG methods.
\end{itemize}

\section{Related Work}
\subsection{Single-Source Domain Generalization}
The goal of SDG is to strengthen DL models to achieve consistent and accurate performance across target domains, even when training relies solely on data from a single source. In addition, SDG is practical in real-world scenarios, where available data from unseen target domains are restricted or nonexistent.

Recent SDG studies have been dedicated to improving model robustness and generalizability in innovative augmentation techniques. Conventionally, Cutout \citep{devries2017improved} enforced the robustness of the model to corruption by deliberately occluding via random patch masking from training images. \citet{xu2020robust} proposed RandConv, which transforms image intensity and texture using a randomly initialized convolutional layer. Apart from the data-level augmentation, MixStyle \citep{zhou2021domain} and RSC \citep{huang2020self} deal with feature representation in latent space. MixStyle synthesizes new domain characteristics by blending styles of instance-level feature statistics, whereas RSC iteratively eliminates dominant features that induce large gradients to learn more general features. 

In medical imaging, \citet{chen2020realistic} proposed AdvBias, which employs adversarial perturbation-based augmentation to produce plausible, realistic signal corruption for intensity inhomogeneities. Moreover, CSDG \citep{ouyang2022causality} takes a unique approach by blending two augmented images via a pseudo-correlation map to reduce spurious spatial correlations that could hinder generalization. \citet{su2023rethinking} explored a novel combination of saliency-balancing fusion with location-scale augmentation (SLAug) that engages the inherent segment information with lower source risk. In contrast to these methods, which directly manipulate the original input, we introduce Fourier-based augmentation to leverage semantic amplitude manipulation and impose structural boundaries from the phase spectrum to boost the domain generalization capabilities specialized for MIS.

\subsection{Fourier-Based Data Manipulation}
Recent studies have explored more advanced augmentation methods, such as manipulating images in the frequency domain. Specifically, they used amplitude manipulation with various strategies, such as swapping \citep{yang2022source, chen2021amplitude, yang2020fda}, masking \citep{hu2022domain, nam2021frequency}, Mixup \citep{liu2021feddg, xu2021fourier, li2023frequency}, and perturbation \citep{zhao2022test, chattopadhyay2023pasta}. 

\citet{yang2020fda} proposed a simple method to generate source images in the target style by replacing the low-frequency amplitude spectrum between source and target domains in swapping and masking-based approaches. In contrast, \citet{chen2021amplitude} designed the amplitude-phase recombination to incorporate the phase component of the source image with the amplitude component of the target image. Additional techniques have emerged, such as Mixup \citep{li2023frequency} and perturbation \citep{liu2021feddg,xu2021fourier}, which involve reconstructing data by linearly interpolating amplitudes or adding noise. \citet{peng2023rain} proposed a novel data augmentation technique called phase Mixup, emphasizing task-relevant objects in interpolations and improving input-level regularization and class consistency for target models. In addition, \citet{xu2023semantic} devised the semantic-aware Mixup (SA-Mixup) for image augmentation by simultaneously manipulating the Fourier component on both amplitude and phase components. 

Nevertheless, these techniques have encountered practical constraints because they depend on the accessibility of the target domain samples or random augmentations with predetermined parameters (\eg, mask or patch sizes) that may not always align with the needs of the model or data. In this context, FIESTA adequately adjusts the amplitude and phase components using semantically measured factors in a single-source domain, circumventing the need for target domain samples or arbitrary parameter settings.

\subsection{Uncertainty in Medical Image Segmentation}
Uncertainty estimation plays a crucial role in a variety of MIS applications, serving as an essential tool for model regularization. \citet{bateson2022source} and \citet{wu2023upl} implemented this concept by adding an objective function to minimize uncertainty loss, which was calculated using Shannon entropy \citep{lin1991divergence}. Further, \citet{ijcai2022p201} introduced an innovative uncertainty-guided contrastive learning mechanism, building upon this approach. This method employs the uncertainty map generated for each unlabeled image to compute the contrastive loss, decreasing the incidence of incorrect sample identification by applying uncertainty-based guidance. As a learning strategy, \citet{shin2023sdc} designed a method over uncertainty-regularized pseudo-label refinement exploiting the uncertainty, generating pseudo-labels for unlabeled target domain data for self-training. \citet{gaillochet2022taal} devised an innovative semi-supervised active learning method that capitalizes on uncertainty information extracted via data augmentation for more informed sample selection in the training and testing phases.

Diverging from these conventional methods, FIESTA applies estimated uncertainty as a pivotal source of information for data augmentation. The proposed method allows for a more focused approach, ensuring that the perspectives are varied in the most confusing area for segmentation. Consequently, uncertainty-guided data augmentation can reinforce learning in areas prone to ambiguity.

\section{Method}
The FIESTA approach enhances the generalizability and segmentation performance of the model across cross-domain scenarios by facilitating data augmentation through semantic manipulation in the frequency domain and UG using epistemic uncertainty estimation. As illustrated in Fig. \ref{fig:FIESTA-framework}, FIESTA comprises three principal steps: (i) context-aware augmentation (Section \ref{sec:ca-aug}), (ii) location-aware augmentation (Section \ref{sec:la-aug}), and (iii) uncertainty-guided mutual augmentation (Section \ref{sec:ug-aug}). Empirically, the FAT was employed for context-aware and location-aware augmentation (Section \ref{sec:fat}). In context-aware augmentation, FAT was applied to adjust the overall image context, resulting in globally augmented images that maintain structural coherence and encourage contextually enriched representation. Preliminary processing was conducted using a spatial transformation function before the FAT execution in location-aware augmentation to further enrich the distinctive styles in each segment location. Uncertainty-guided mutual augmentation was performed to boost segmentation capabilities from the indiscernible or ambiguous parcellated areas using two augmentation images from different perspectives and estimating their uncertainty.

\subsection{Preliminary: Fourier Transformation}
The motivation behind FIESTA is introducing an advanced form of data augmentation beyond traditional spatial transformations, engaging with the frequency components of medical images. An effective strategy to create augmented images that retain the essential characteristics of the original data while introducing variations in and exploitation of the amplitude and phase spectrum is to use the Fourier transform \citep{yang2020fda,chen2021amplitude,chattopadhyay2023pasta}. Prior to discussing the specifics, we revisit the formulation of the Fourier transform and cover the necessary preliminaries on extracting amplitude and phase spectrum from an image in the spatial domain, accordingly.

Given a grayscale two-dimensional (2D) image $\x$ with $H\times W$ resolution, the Fourier transform over the input $\x$ can be formulated as follows:
\begin{equation}
    \mathcal{F}(\x)(u,v)=\sum_{h=0}^{H-1} \sum_{w=0}^{W-1} \x(h, w) e^{-j 2 \pi\left(\frac{h}{H} u+\frac{w}{W} v\right)},
\end{equation}
where $\mathcal{F}(\cdot)$ denotes the 2D fast Fourier transform (FFT) \citep{nussbaumer1982fast} that transforms the image of the spatial domain into the frequency domain, and $u$ and $v$ indicate the coordinates in the frequency space. The amplitude $\mathcal{A}(\x)\in\mathbb{R}^{U\times V\times 1}$ and phase $\mathcal{P}(\x)\in\mathbb{R}^{U\times V\times 1}$ components are computed from input $\x$ as follows:
\begin{equation}
    \mathcal{A}(\x)(u, v)=\sqrt{R^2(\x)(u, v)+I^2(\x)(u, v)},\label{eq:amp}
\end{equation}
\begin{equation}
    \mathcal{P}(\x)(u, v)=\arctan \left[\frac{I(\x)(u, v)}{R(\x)(u, v)}\right],\label{eq:pha}
\end{equation}
where $R(\x)$ and $I(\x)$ represent the real and imaginary parts of the Fourier transform, respectively. Without loss of generality, the abbreviations of amplitude $\mathcal{A}(\x)(u,v)$ and phase $\mathcal{P}(\x)(u,v)$ are $\mathcal{A}(\x)$ and $\mathcal{P}(\x)$, respectively. The inverse FFT (iFFT), represented as $\mathcal{F}^{-1}(\cdot)$, is calculated to translate spectral signals involving the amplitude and phase from the frequency domain back into the spatial domain. For convenience, the FFT $\mathcal{F}(\cdot)$ and iFFT $\mathcal{F}^{-1}(\cdot)$ for the remaining sections employ a shift operator that multiplies $(-1)^{u+v}$ by the amplitude and phase spectrum to shift the low-frequency components to the center.

\subsection{Problem Definition}
First, we let $\mathcal{D}^S$ and $\mathcal{D}^T$ denote the source and target domains, respectively. Moreover, $\{\x^S_{i}, \y^S_{i}\}_{i=1}^{N_S}$ represents the set of $N_S$ training images and their corresponding ground truths in the source domain $\mathcal{D}^S$, and $\{\x^T_{i}, \y^T_{i}\}_{i=1}^{N_T}$ represents $N_T$ unseen test images and their corresponding ground truths in the target domain $\mathcal{D}^T$, where $\x^S, \x^T\in\mathbb{R}^{W\times H\times 1}$, $\y^S, \y^T\in\{0,1\}^{W\times H\times C}$. In addition, $W$, $H$, and $C$ indicate the width, height, and number of classes (\ie, parcellated organs), respectively. For the MIS task, the segmentation model $\s_\psi$ parameterized by $\psi$ is typically trained in the source domain using fully supervised learning as follows:
\begin{equation}
    \psi^* := \argmin_{\psi}\frac{1}{N_S}\sum^{N_S}_{i=1}\mathcal{L}\left(\s_\psi(\x_i^S), \y_i^S\right),
\end{equation}
where $\mathcal{L}$ denotes a type of segmentation loss, and $\psi^*$ indicates the optimized weights of the segmentation model $\s_\psi$.

In the SDG scenario, where the target domain remains unseen during the training phase, FIESTA aims to enhance the segmentation performance on such unseen target domains $\mathcal{D}^T$. To accomplish this, the role of FIESTA is to effectively augment the source image to encourage learning so that the segmentation model is robust and sufficiently stable for other unseen domains as follows:
\begin{equation}
    \psi^* := \argmin_{\psi}\frac{1}{N_S}\sum^{N_S}_{i=1}\mathcal{L}\left(\s_\psi(\operatorname{Aug}(\x_i^S)), \y_i^S\right),
\end{equation}
where $\operatorname{Aug}(\cdot)$ denotes each augmentation method in FIESTA, including context-aware augmentation, location-aware augmentation, and uncertainty-guided mutual augmentation. This strategy is achieved by extending the source domain's data distribution in a way that anticipates and compensates for potential discrepancies between the source and target domains, all while not having direct access to the target domain dataset $\{\x^T_{i}, \y^T_{i}\}_{i=1}^{N_T}$. For brevity, we substituted the notation for the source dataset $\{\x^S_{i}, \y^S_{i}\}_{i=1}^{N_S}$ with $\{\x_{i}, \y_{i}\}_{i=1}^{N}$ hereafter.

\begin{figure}[t]
    \centering
    \includegraphics[width=1\linewidth]{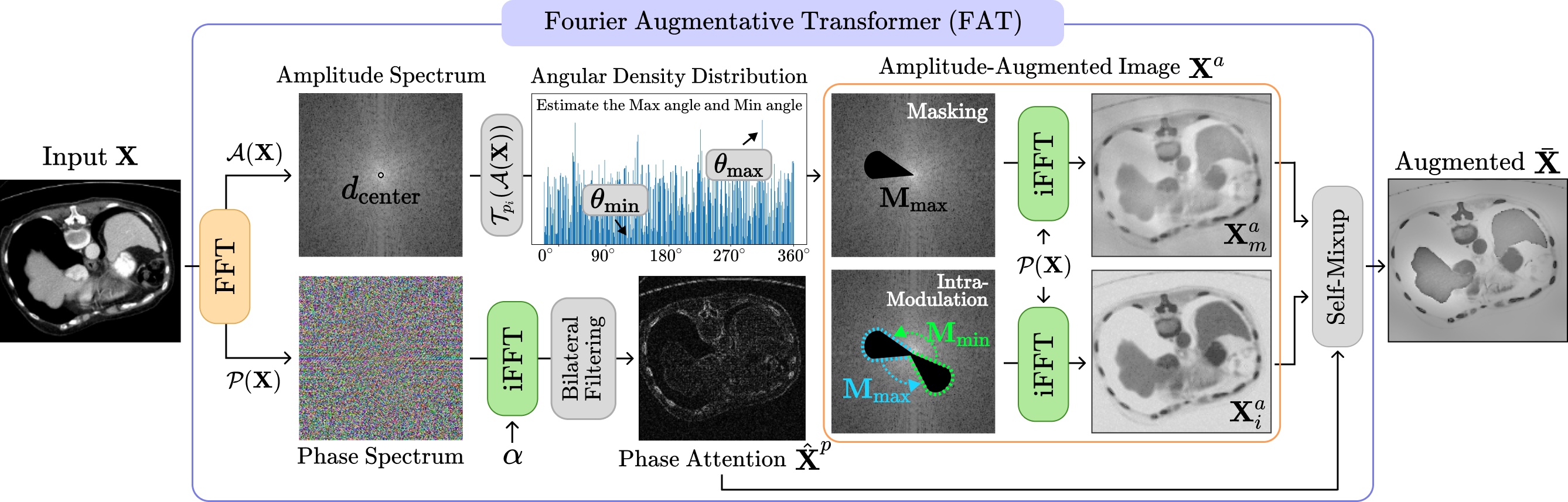}
    \caption{Overall Fourier augmentative transformer (FAT) framework, applying novel masking and intra-modulation techniques to the amplitude spectrum with phase attention via an advanced filtering strategy (FFT: fast Fourier transform, iFFT: inverse FFT).}
    \label{fig:fat-framework}
\end{figure}

\subsection{Fourier Augmentative Transformer}\label{sec:fat}
Given the amplitude component $\mathcal{A}(\x)$ and phase component $\mathcal{P}(\x)$ through Eqs. \eqref{eq:amp} and \eqref{eq:pha}, two types of manipulation were performed on each Fourier component: (1) amplitude masking and intra-modulation and (2) phase attention. The inner working of FAT is illustrated in Fig. \ref{fig:fat-framework}.

\subsubsection{Amplitude Masking and Intra-Modulation}
Enhancing the ability of the model to adapt to image textures and contrast variations is crucial in medical imaging, particularly in differentiating tissue types. To facilitate this, we have developed amplitude masking and intra-modulation techniques, as manipulating the amplitude can significantly alter domain-variant global contrasts of images within the spatial domain, provoking their textural differences \citep{yang2022source,li2023frequency}. Thus, the amplitude transformation $\mathcal{T}{_{p_i}}(\cdot)$ is employed, which reverses the scale of the amplitude component according to a randomly produced probability $p_i$ as follows:
\begin{equation}\label{eq:amp-trans}
    \mathcal{T}{_{p_i}}(\mathcal{A}(\x))=\left\{\begin{array}{ll}
    |\operatorname{med}(\mathcal{A}(\x))-\mathcal{A}(\x)| & \text { if  } p_i \geq 0.5\\
    \mathcal{A}(\x) & \text { otherwise, }
    \end{array}\right.
\end{equation}
where $|\cdot|$ and $\operatorname{med}(\mathcal{A}(\x))$ denote the absolute operation and median value of the amplitude $\mathcal{A}(\x)$, respectively. This process produces various aspects of image content because the amplitude transformation influences the change in domain-variant characteristics in the spatial domain. Through these capabilities, the amplitude transformation can make the model less sensitive to unknown textures in unseen domains. Afterward, we hypothesize that areas with a high amplitude magnitude at specific angular points are significant because they engage abundant domain-specific information. Such angular points represent where domain-variant features are most marked, such as superficial patterns and fine details distributed across the image, allowing crucial insight into the structural and textural elements of an image. To estimate meaningful angular points, we measured the angular density distribution $f(\cdot)$, which reveals the information amount for each angle $\theta$ from 0$^\circ$ to 359$^\circ$ based on the center coordinate $d_{\text{center}}$ in the transformed amplitude $\bar{\mathcal{A}}(\x)=\mathcal{T}{_{p_i}}({\mathcal{A}}(\x))$:
\begin{equation}
    f\left(\theta; \bar{\mathcal{A}}(\x)\right)=\sum^{R}_{r=1}\bar{\mathcal{A}}(\x)_{\lfloor r \cos \theta\rfloor,\lfloor r \sin \theta\rfloor},\label{eq:ang-den-dis}
\end{equation}
where $R$ and $\lfloor\cdot\rfloor$ indicate the width or height of $\bar{\mathcal{A}}(\x)$ and the floor function, respectively. In this way, we derived the angles with the maximum and minimum densities as $\theta_{\text{max}}, \theta_{\text{min}}\sim f(\theta; \bar{\mathcal{A}}(\x))$ under the measured angular density distribution (Fig. \ref{fig:fat-framework}). Empirically, $\theta_{\text{max}}$ represents the direction involving distinct features or edges in the image, whereas $\theta_{\text{min}}$ is a relatively less pronounced direction. With these estimated angles, we created a partial sector mask\footnote{The partial sector shape is adopted as the mask form, as the pattern structure of the amplitude spectrum gradually expands outward from the center frequency.} $\mathbf{M}\in\mathbb{R}^{U\times V\times 1}$ that masks the amplitude region within the angle $\bar{\theta}$ with the length of the radius $\bar{r}$ from the center coordinate $d_{\text{center}}$ of the amplitude spectrum as follows:
\begin{equation}\label{eq:amp-masking}
    \mathbf{M}(\theta, k, r)=\left\{\begin{array}{ll}
    1 & \text { if  } \bar{r}\in\left[d_{\text{center}},d_{\text{center}}+r\right] \text{ and } \bar{\theta}\in\left[\theta - \frac{k}{2},\theta + \frac{k}{2}\right]\\
    0 & \text { otherwise, }
    \end{array}\right.
\end{equation}
where $0<r<\frac{\operatorname{min}\{U,V\}}{2}$ and $k$ denotes randomly chosen angle. Thus, we can obtain the maximum and minimum angle masks $\mathbf{M}_{\text{max}}=\mathbf{M}(\theta_{\text{max}},k, r)$ and $\mathbf{M}_{\text{min}}=\mathbf{M}(\theta_{\text{min}},k, r)$. The produced partial sector masks denote the indicator function conducted in the polar coordinate system.

For amplitude masking, we applied the maximum angle mask $\mathbf{M}_{\text{max}}$ that could diminish the most dominant directional attributes, making other textural characteristics more prominent. By multiplying the amplitude $\bar{\mathcal{A}}(\x)$ by the given maximum angle mask $\mathbf{M}_{\text{max}}$ and applying the iFFT $\mathcal{F}^{-1}$ with a phase component $\mathcal{P}(\x)$, the amplitude-masked image $\x^a_m$ is generated as follows:
\begin{equation}
    \x^a_m:=\mathcal{F}^{-1}\left(\bar{\mathcal{A}}(\x)\odot(1-\mathbf{M}_{\text{max}}), \mathcal{P}(\x)\right),\label{eq:amp-masked-img}
\end{equation}
where $\odot$ denotes the Hadamard product.

To perform the amplitude intra-modulation, we replaced the values in amplitude $\bar{\mathcal{A}}(\x)$ using the generated maximum angle mask $\mathbf{M}_{\text{max}}$ and minimum angle mask $\mathbf{M}_{\text{min}}$. Altering the primary directional information between these angle masks maintains the overall patterns and structures but emphasizes or de-emphasizes intrinsic properties for certain directions. This process introduces new data variations, assisting models in learning to generalize across scenarios or imaging modalities better. For this objective, we swapped values in $\mathbf{M}_{\text{max}}$ with the corresponding positions $(u,v)$ in $\mathbf{M}_{\text{min}}$, and vice versa:
\begin{equation}\label{eq:amp-intra-mod}
    \bar{\mathcal{A}}'(\x)=\left\{\begin{array}{ll}
    \bar{\mathcal{A}}(\x)(u_1,v_1)[\bar{\mathcal{A}}(\x)\odot\mathbf{M}_{\text{max}}] & \text { if  } (u_1,v_1)\in\mathbf{M}_{\text{min}}\\
    \bar{\mathcal{A}}(\x)(u_2,v_2)[\bar{\mathcal{A}}(\x)\odot\mathbf{M}_{\text{min}}] & \text { if  } (u_2,v_2)\in\mathbf{M}_{\text{max}}\\
    \bar{\mathcal{A}}(\x) & \text { otherwise. }
    \end{array}\right.
\end{equation}
Subsequently, we obtained an intra-modulated image $\x^a_i$ based on the iFFT $\mathcal{F}^{-1}$ using $\bar{\mathcal{A}}'(\x)$ as follows:
\begin{equation}\label{eq:amp-intra-mod-img}
    \x^a_i:=\mathcal{F}^{-1}\left(\bar{\mathcal{A}}'(\x), \mathcal{P}(\x)\right).
\end{equation}

\subsubsection{Phase Attention Mechanism}
Harnessing phase information could reflect the shape and structural integrity of objects (\ie, domain-invariant properties), making the model more susceptible to various types of object morphology \citep{liu2021feddg,peng2023rain}. To enforce high-level semantics in the amplitude-manipulated images $\x^a_m$ and $\x^a_i$, we devised a phase-based attention mechanism for data augmentation. The phase component was first reconstructed into the spatial domain image ${\x}^p = \mathcal{F}^{-1}(\alpha, \mathcal{P}(\x))$, where $\alpha$ is a constant representing the amplitude component to conduct the iFFT. Given the reconstructed phase image ${\x}^p$, we refine this image by exploiting the bilateral filter \citep{tomasi1998bilateral} that maintains anatomic appearance, such as tissue boundaries, while smoothing out the noise. Bilateral filtering considers the spatial proximity (using the spatial kernel $G_{\sigma_s}$) and the intensity difference (using the range kernel $G_{\sigma_r}$) between pixels. Each image pixel is replaced with a weighted average of its neighbors. The phase attention applying bilateral filtering underlying these favorable properties is defined as follows:

\begin{equation}
    \hat{\x}^p(i,j) = \frac{\sum_{(i',j')\in \Omega}G_{\sigma_s}(||(i',j')-(i, j)||)\cdot G_{\sigma_r}(\|{\x}^p(i',j')-{\x}^p(i, j)\|)\cdot{\x}^p(i',j')}{\sum_{(i',j')\in \Omega}G_{\sigma_s}(||(i',j')-(i, j)||)\cdot G_{\sigma_r}(\|{\x}^p(i',j')-{\x}^p(i, j)\|)},\label{eq:bilateral}
\end{equation}
where $(i,j)$ and ${\x}^p(i,j)$ denote the center pixel location being processed and the pixel intensity, respectively, in the reconstructed phase image ${\x}^p$ at position $(i,j)$. In addition, $\Omega$ is the window (\ie, filter mask) around pixel $(i,j)$, and $\|\cdot\|$ is the $\ell_2$-norm to calculate the Euclidean distance from $(i,j)$ to $(i', j')$. The weight depends on the spatial distance and intensity difference, allowing the filter to highlight edges where the intensity significantly changes.

\subsubsection{Self-Mixup Strategy}
Eventually, Fourier-augmented image $\bar{\x}$ from the FAT is acquired by incorporating $\x^a_m$, $\x^a_i$, and $\hat{\x}^p$ based on the self-Mixup strategy, as follows:
\begin{equation}
    \bar{\x} := \lambda\cdot(\x^a_m\odot\hat{\x}^p) + (1-\lambda)\cdot(\x^a_i\odot\hat{\x}^p),\label{eq:fat}
\end{equation}
where the hyperparameter $\lambda$ controls the calibration strength for self-Mixup. By properly enforcing domain-invariant phenotypes via phase attention, the model can recognize structural integrity and perform precise segmentation. Algorithm \ref{pseudocode2} provides the pseudocode for the overall procedures of the FAT.

\begin{algorithm}[!t]
\footnotesize
\caption{Pseudo algorithm for the FAT} 
    \begin{algorithmic}[1]\label{pseudocode2}
    \REQUIRE{Original input $\x$, FFT $\mathcal{F}(\cdot)$, iFFT $\mathcal{F}^{-1}(\cdot)$, Amplitude transformation $\mathcal{T}_{p_i}(\cdot)$, Angular density distribution $f(\cdot)$, Intra-modulation $\operatorname{IM}(\cdot)$, Constant value $\alpha$, Bilateral filter $\operatorname{BF}(\cdot)$, Randomly selected angle $k$, Randomly selected radius $r$, Weighting coefficient $\lambda$.}\\
    
    \STATE $\mathcal{A}(\x), \mathcal{P}(\x)$ $\leftarrow$ $\mathcal{F}(\x)$ \hfill{$\triangleright$ Eqs. \eqref{eq:amp} and \eqref{eq:pha}}
    \STATE $\bar{\mathcal{A}}(\x) = \mathcal{T}_{p_i}(\mathcal{A}(\x))$
    \hfill{$\triangleright$ Eq. \eqref{eq:amp-trans}}
    \STATE $\theta_{\text{max}}, \theta_{\text{min}}\sim f(\theta; \bar{\mathcal{A}}(\x))$
    \hfill{$\triangleright$ Eq. \eqref{eq:ang-den-dis}}
    \STATE $\mathbf{M}_{\text{max}}$, $\mathbf{M}_{\text{min}}$ $\leftarrow$ $\mathbf{M}(\theta_{\text{max}},k, r)$, $\mathbf{M}(\theta_{\text{min}},k, r)$
    \hfill{$\triangleright$ Eq. \eqref{eq:amp-masking}}
    
    \STATE{\em // Amplitude Masking}
    \STATE $\x^a_m:=\mathcal{F}^{-1}\left(\bar{\mathcal{A}}(\x)\odot(1-\mathbf{M}_{\text{max}}), \mathcal{P}(\x)\right)$
    \hfill{$\triangleright$ Eq. \eqref{eq:amp-masked-img}}
    
    \STATE{\em // Intra-Modulation}
    \STATE $\bar{\mathcal{A}}'(\x) = \operatorname{IM}(\bar{\mathcal{A}}(\x), \mathbf{M}_{\text{max}}, \mathbf{M}_{\text{min}})$
    \hfill{$\triangleright$ Eq. \eqref{eq:amp-intra-mod}}
    \STATE $\x^a_i:=\mathcal{F}^{-1}\left(\bar{\mathcal{A}}'(\x), \mathcal{P}(\x)\right)$
    \hfill{$\triangleright$ Eq. \eqref{eq:amp-intra-mod-img}}
    
    \STATE{\em // Phase Attention Mechanism}
    \STATE ${\x}^p = \mathcal{F}^{-1}(\alpha, \mathcal{P}(\x))$
    \STATE $\hat{\x}^p = \operatorname{BF}(\x^p)$
    \hfill{$\triangleright$ Eq. \eqref{eq:bilateral}}

    \STATE{\em // Self-Mixup Strategy}
    \STATE $\bar{\x} := \lambda\cdot(\x^a_m\odot\hat{\x}^p) + (1-\lambda)\cdot(\x^a_i\odot\hat{\x}^p).$
    \hfill{$\triangleright$ Eq. \eqref{eq:fat}}
    \RETURN{Fourier-augmented image $\bar{\x}$}
    \end{algorithmic}
\end{algorithm}

\subsection{FIESTA}
The objective of FIESTA is to improve segmentation performance and model generalizability across domains using advanced data augmentation techniques. This approach includes three steps: context-aware augmentation, location-aware augmentation, and uncertainty-guided mutual augmentation. Context-aware augmentation modifies the entire image context, whereas location-aware augmentation tailors adjustments to specific segment locations. Both augmentation techniques are integrated through uncertainty-guided mutual augmentation, exploiting epistemic uncertainty to refine the segmentation of ambiguous areas, enhancing the adaptability of the model to diverse imaging scenarios.

\subsubsection{Context-Aware Augmentation}\label{sec:ca-aug}
Context-aware augmentation is performed via the proposed FAT with the original input $\x$ such that $\bar{\x}_{\text{CA}}=\operatorname{FAT}(\x)$. The reason for this context-aware augmentation strategy is to create variations that affect the overall image context. Particularly, the FAT introduces perturbations to vital context features via amplitude masking and intra-modulation, whereas phase attention maintains semantic information (Eq. \eqref{eq:fat}). This delicate balance via self-Mixup means that the augmented images take into account the intrinsic characteristics of the original images, although certain features were altered. By exposing the model to numerous context changes using this method, the model becomes better equipped to address unseen images with diverse contextual attributes effectively. As such, the segmentor $\s_\psi$ using context-aware FAT (C-FAT) images $\bar{\x}_{\text{CA}}$ was optimized as the following objective function:
\begin{equation}
    \mathcal{L}_{\text{CA}}=\mathcal{L}_{\text{ce}}(\y,\bar{\y}_{\text{CA}})+\mathcal{L}_{\text{dice}}(\y,\bar{\y}_{\text{CA}}),\label{eq:loss-ca}
\end{equation}
where $\bar{\y}_{\text{CA}} = \s_\psi(\bar{\x}_{\text{CA}})$ and $\y$ denotes the ground truth. To optimize the segmentor $\s_\psi$, we applied a combination of cross-entropy loss $\mathcal{L}_{\text{ce}}$ and Dice loss $\mathcal{L}_{\text{dice}}$, prevalent objective functions in image segmentation.

\begin{figure}[t]
    \centering
    \includegraphics[width=1\linewidth]{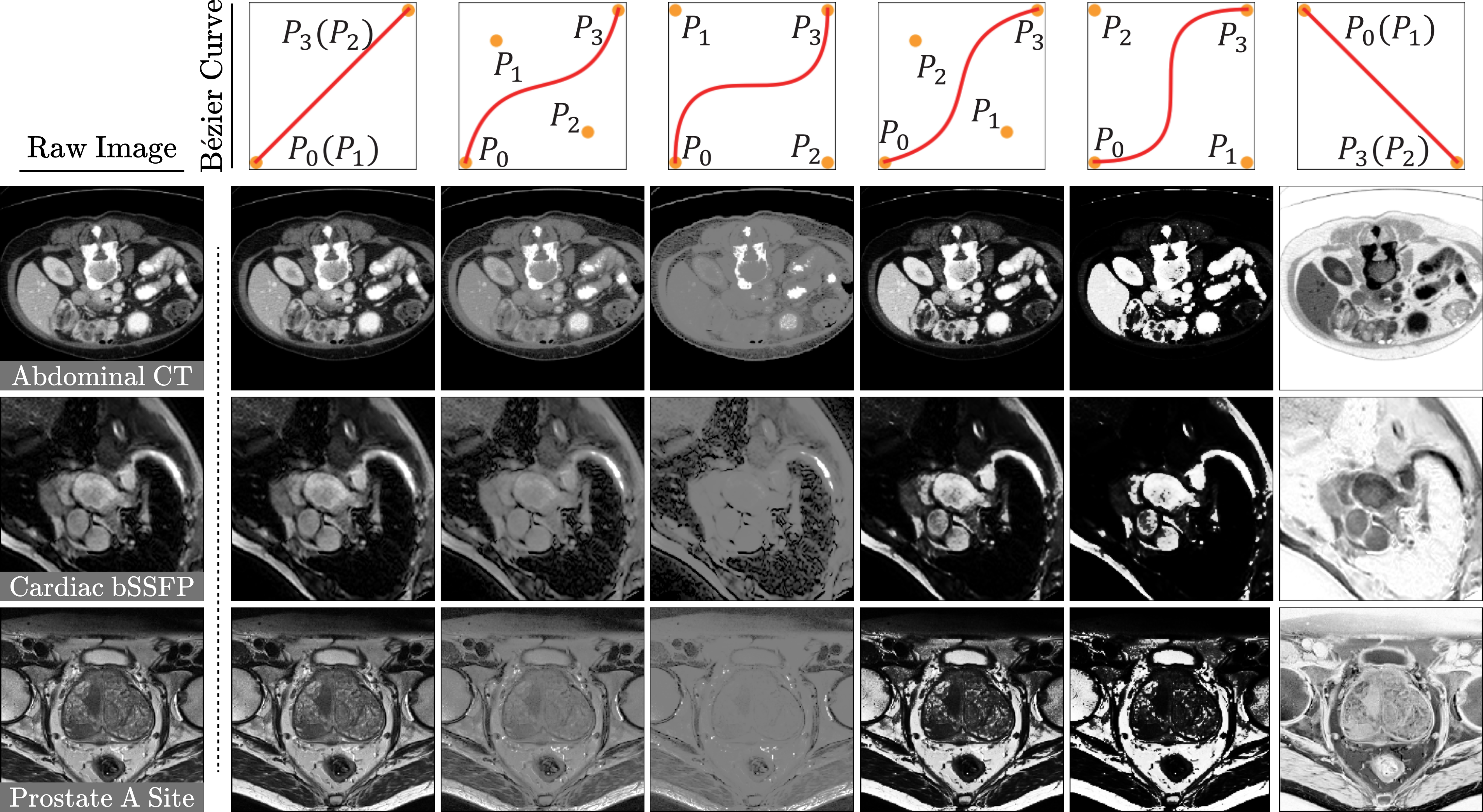}
    \caption{Examples of transformed raw images according to the B$\Acute{e}$zier curve variant.}
    \label{fig:bezier-res}
\end{figure}

\subsubsection{Location-Aware Augmentation}\label{sec:la-aug}
The purpose of location-aware augmentation is to simulate any segment-wise contrast inconsistency that may be caused by unseen domain modalities. Motivated by Model Genesis \citep{zhou2019models}, we applied the smooth and monotonic B$\Acute{e}$zier curve \citep{mortenson1999mathematics} as a non-linear transformation function to adjust the contrast in each segment region before applying the FAT. In Fig. \ref{fig:bezier-res}, the transformation function is generated using two starting and ending points of $P_0=(v_{\text{min}},v_{\text{min}})$ and $P_3=(v_{\text{max}},v_{\text{max}})$ through the intensity scale of the input (\eg, the min-max normalized image: $v_{\text{min}}=0$ and $v_{\text{max}}=1$) to constrain the value range. In addition, two control points of $P_1$ and $P_2$ are randomly generated values from a limited range $[v_{\text{min}}, v_{\text{max}}]$:
\begin{equation}
    B\Acute{e}izer(t) = \sum^n_{i=0}P_i(1-t)^{n-i}t^i,\enspace n=3, t\in[v_{\text{min}},v_{\text{max}}],
\end{equation}
where $t$ denotes a fractional value along the line length. For brevity, we defined the B$\Acute{e}$zier curve-based transformation function as $\mathcal{B}_{p}(\cdot)$, where $p$ is the probability of whether the inverse transformation is executed for the region of each class label $\y_c$ (\ie, the segment region including the background). Activating the inverse transformation in $\mathcal{B}_{p}$ ($p>0.5$) means that the starting and ending point ($P_0$ and $P_3$) values are switched (\ie, $P_0=(v_{\text{min}},v_{\text{max}})$ and $P_3=(v_{\text{max}},v_{\text{min}})$), as illustrated in Fig. \ref{fig:bezier-res} on the right. For the expansion of data diversity, we further applied the linear scaling and shifting terms inspired by \citet{su2023rethinking}, such that the location-aware FAT (L-FAT) image $\bar{\x}_{\text{LA}}$ was obtained after the FAT operation:
\begin{equation}
    \bar{\x}_{\text{LA}} = \operatorname{FAT}\left(\sum^C_{c=1}\alpha_c\cdot\mathcal{B}_{p_c}(\x\odot\y_c) + \beta_c\right), \alpha_c\sim\mathcal{TN}(1,\sigma_1), \beta_c\sim\mathcal{TN}(0,\sigma_2),\label{eq:fat-bezier}
\end{equation}
where $\sigma_1$ and $\sigma_2$ represent the standard deviations of the two truncated Gaussian distributions $\mathcal{TN}(\cdot)$, and $C$ denotes the number of classes. Similar to the optimization of context-aware augmentation, the segmentor $\s_\psi$ was trained using $\bar{\y}_{\text{LA}} = \s_\psi(\bar{\x}_{\text{LA}})$ based on the losses of $\mathcal{L}_{\text{ce}}$ and $\mathcal{L}_{\text{dice}}$ as follows:
\begin{equation}
    \mathcal{L}_{\text{LA}}=\mathcal{L}_{\text{ce}}(\y,\bar{\y}_{\text{LA}})+\mathcal{L}_{\text{dice}}(\y,\bar{\y}_{\text{LA}}).\label{eq:loss-la}
\end{equation}

\subsubsection{Uncertainty-Guided Mutual Augmentation}\label{sec:ug-aug}
The primary idea of harmonizing uncertainty in the augmentation process is to train the segmentor by focusing on challenging areas that fail segmentation owing to the similarity of contrasts between or within organs \citep{su2023rethinking}. To this end, the epistemic uncertainty is derived using the segmentor $\s_\psi$ learned from each step of the two augmentation methods (Sections \ref{sec:ca-aug} and \ref{sec:la-aug}). To estimate the uncertainty $\mathbf{U}$, we adopted the Shannon entropy \citep{lin1991divergence} $\mathbf{U} = -\sum^C_{c=1}p_c\log_2(p_c)$, where $p_c$ is the softmax activated logits of the segmentor for each segment $c$ (\ie, the posterior probability). This approach can yield the uncertainty maps of C-FAT $\bar{\x}_{\text{CA}}$ and L-FAT $\bar{\x}_{\text{LA}}$ from the trained segmentor $\s_\psi$: $\mathbf{U}_c=\s_\psi(\bar{\x}_{\text{CA}})$ and $\mathbf{U}_l=\s_\psi(\bar{\x}_{\text{LA}})$. These maps are adequately fused via aggregative fusion that considers the heterogeneity between $\mathbf{U}_c$ and $\mathbf{U}_l$ to leverage such uncertainty maps as guidance to induce augmentation over confound areas. Thus, UG $\mathbf{U}_{\text{map}}$ is created as follows:
\begin{equation}
    \mathbf{U}_{\text{map}} = \operatorname{GaussianBlur}\left(\frac{1}{2}\left(\max\{\mathbf{U}_c, \mathbf{U}_l\} + \frac{\mathbf{U}_c + \mathbf{U}_l}{2}\right)\right),\label{eq:umap}
\end{equation}
where $\operatorname{GaussianBlur}$ denotes the smoothing function, applying a Gaussian filter to the uncertainty map. The first term in Eq. \eqref{eq:umap} highlights areas where either one of the augmentation processes has discerned substantial ambiguity, offering that these regions are prioritized in the subsequent augmentation guidance. By contrast, the second term in Eq. \eqref{eq:umap} provides a comprehensive view that aligns with the uncertainty of both methods while smoothing out the extremities to prevent overemphasis on discrepancies. With such advantages, the mutually augmented (MA) image $\bar{\x}_{\text{MA}}$ was produced using C-FAT $\bar{\x}_{\text{CA}}$ and L-FAT $\bar{\x}_{\text{LA}}$ by leveraging a variant over the linearly interpolated fusion method in \citet{ouyang2022causality} and \citet{su2023rethinking}. Specifically, we multiplied the UG $\mathbf{U}_{\text{map}}$ and its reverse by the $\bar{\x}_{\text{CA}}$ and $\bar{\x}_{\text{LA}}$ as follows:

\begin{algorithm}[!t]
\footnotesize
\caption{Pseudo algorithm for the FIESTA}
    \begin{algorithmic}[1]\label{pseudocode}
    \REQUIRE{Training dataset $\{\x, \y\}$, segmentation network $\s_\psi$, Tunable parameters $\psi_*$, Fourier augmentative transformer $\operatorname{FAT}(\cdot)$, B$\Acute{e}$zier transformation $\mathcal{B}(\cdot)$, aggregative fusion $\operatorname{AF}(\cdot)$ Cross-entropy loss $\operatorname{CE}(\cdot)$, Dice loss $\operatorname{Dice}(\cdot)$, Number of classes $C$, posterior probability $p$, Learning rate $\eta$.}
    
    \FOR{$epoch = 1, 2,\dots$}
    \STATE Randomly select $B$ mini-batches from all training samples 
    \FOR{$b\leftarrow1$ to $B$}

    \STATE{\em // Step 1. Context-aware augmentation using FAT}
    \STATE $\bar{\x}_{\text{CA}} = \operatorname{FAT}(\x)$ \hfill{$\triangleright$ Eq. \eqref{eq:fat}}
    \STATE $\bar{\y}_{\text{CA}} = \s_{\psi_1}(\bar{\x}_{\text{CA}})$

    \STATE Update the segmentation network $\s_{\psi_1}$ under $\bar{\y}_{\text{CA}}$:
    \STATE $\psi_2 \leftarrow \psi_1 - \eta\nabla_{\psi_1}(\operatorname{CE}(\y, \bar{\y}_{\text{CA}}) + \operatorname{Dice}(\y, \bar{\y}_{\text{CA}}))$ \hfill{$\triangleright$ Eq. \eqref{eq:loss-ca}}
    \STATE Calculate uncertainty $\mathbf{U}_c=-\sum^C_{c=1}p_c\log_2(p_c)$, where $p=\s_{\psi_2}(\bar{\x}_{\text{CA}})$
    
    \STATE{\em // Step 2. Location-aware augmentation with FAT}
    \STATE $\bar{\x}_{\text{LA}} = \operatorname{FAT}(\mathcal{B}(\x,\y))$ \hfill{$\triangleright$ Eq. \eqref{eq:fat-bezier}}
    \STATE $\bar{\y}_{\text{LA}} = \s_{\psi_2}(\bar{\x}_{\text{LA}})$
    \STATE Update the segmentation network $\s_{\psi_2}$ under $\bar{\y}_{\text{LA}}$: 
    \STATE $\psi_3 \leftarrow \psi_2 - \eta\nabla_{\psi_2}(\operatorname{CE}(\y, \bar{\y}_{\text{LA}}) + \operatorname{Dice}(\y, \bar{\y}_{\text{LA}}))$ \hfill{$\triangleright$ Eq. \eqref{eq:loss-la}}
    \STATE Calculate uncertainty $\mathbf{U}_l=-\sum^C_{c=1}p_c\log_2(p_c)$, where $p=\s_{\psi_3}(\bar{\x}_{\text{LA}})$

    \STATE{\em // Step 3. Mutual augmentation using uncertainty guidance}
    \STATE $\mathbf{U}_{\text{map}} = \operatorname{AF}(\mathbf{U}_c, \mathbf{U}_l)$ \hfill{$\triangleright$ Eq. \eqref{eq:umap}}
    \STATE $\bar{\x}_{\text{MA}}=\bar{\x}_{\text{CA}}\odot\mathbf{U}_{\text{map}}+\bar{\x}_{\text{LA}}\odot(1-\mathbf{U}_{\text{map}})$ \hfill{$\triangleright$ Eq. \eqref{eq:mutual-aug}}
    \STATE $\bar{\y}_{\text{MA}} = \s_{\psi_3}(\bar{\x}_{\text{MA}})$
    \STATE Update the segmentation network $\s_{\psi_3}$ under $\bar{\y}_{\text{MA}}$:
    \STATE $\psi \leftarrow \psi_3 - \eta\nabla_{\psi_3}(\operatorname{CE}(\y, \bar{\y}_{\text{MA}}) + \operatorname{Dice}(\y,\bar{\y}_{\text{MA}}))$ \hfill{$\triangleright$ Eq. \eqref{eq:loss-ma}}
    \ENDFOR 
    \ENDFOR 
    \RETURN{Trained segmentation network $\s_\psi$}
    \end{algorithmic}
\end{algorithm}

\begin{equation}
    \bar{\x}_{\text{MA}}=\left\{\begin{array}{ll}
    \bar{\x}_{\text{CA}}\odot\mathbf{U}_{\text{map}} + \bar{\x}_{\text{LA}}\odot(1-\mathbf{U}_{\text{map}}) & \text { if  } p_u<0.5\\
    \bar{\x}_{\text{CA}} + \bar{\x}_{\text{LA}} & \text { otherwise, }
    \end{array}\right.\label{eq:mutual-aug}
\end{equation}
where $p_u$ denotes a constraint probability, which is the average of the uncertainty probabilities for the areas corresponding to the foreground from $\mathbf{U}_{\text{map}}$. Using $p_u$ ensures that the mutual augmentation process is controlled to prevent excessive augmentation from being induced where the learning of the model is not yet sufficiently established (\ie, lower model confidence). Hence, the segmentor $\s_\psi$ is optimized using the predicted mask $\bar{\y}_{\text{MA}} = \s_\psi(\bar{\x}_{\text{MA}})$ using $\mathcal{L}_{\text{ce}}$ and $\mathcal{L}_{\text{dice}}$ as follows:
\begin{equation}
    \mathcal{L}_{\text{MA}}=\mathcal{L}_{\text{ce}}(\y,\bar{\y}_{\text{MA}})+\mathcal{L}_{\text{dice}}(\y,\bar{\y}_{\text{MA}}).\label{eq:loss-ma}
\end{equation}

The three mentioned augmentation steps were sequentially executed in a single iteration. Algorithm \ref{pseudocode} provides the pseudocode for the overall procedures for FIESTA.

\section{Experiments}
\subsection{Datasets and Data Preprocessing}
In this literature, we comprehensively conducted experiments by exploiting three cross-domain scenarios of disparate image modalities to substantiate the domain generalizability and outstanding segmentation precision of FIESTA. 

\begin{itemize}
    \item The abdominal cross-modality consists of a collection of 30 abdominal CT volumes \citep{landman2015miccai} with 20 T2-SPIR MRI volumes \citep{kavur2021chaos} with four segments: the liver, right kidney (R-Kid), left kidney (L-Kid), and spleen.
    \item The cardiac cross-sequence includes 45 volumes each of balanced steady-state free precession (bSSFP) and late gadolinium-enhanced (LGE) MRIs \citep{zhuang2022cardiac}. The label set for the cardiac data comprises three segments: the left ventricle (LV), myocardium (MYO), and right ventricle (RV).
    \item The prostate cross-site comprises T2-weighted MRI across six sites: 30 volumes each from Radboud University Nijmegen Medical Centre (Site A) and Boston Medical Center (Site B), 19 volumes from the Hospital Center Regional University of Dijon-Bourgogne (Site C), 13 volumes from University College London (Site D), and 12 volumes each from Beth Israel Deaconess Medical Center (Site E) and Haukeland University Hospital (Site F). In detail, Sites A and B are from the NCI-ISBI13 dataset \citep{Bloch2015nci}; Site C is from the I2CVB dataset \citep{lemaitre2015computer}; Sites D, E, and F are from the PROMISE12 dataset \citep{litjens2014evaluation}. The NCI-ISBI13 and PROMISE12 datasets involve multiple sites; hence, we separated them identically following~\citet{liu2020shape}. The segments in these datasets are binary labels for the prostate region.
\end{itemize}

For data preprocessing, we followed the instructions by \citet{ouyang2020self}, providing a detailed description of preprocessing procedures along with the available code\footnote{\url{https://github.com/cheng-01037/Causality-Medical-Image-Domain-Generalization}}. These three-dimensional (3D)-format datasets were reformatted into 2D slices after the spatial normalization via the zero mean and unit variance and were uniformly cropped and resized to a standard resolution of 192$\times$192. For the CT images within the abdominal dataset, we applied a window ranging from -275 to 125 Hounsfield values \citep{ouyang2020self}. For all MRIs, we trimmed the values corresponding to the upper 0.5$\%$ of the intensity histogram using a clipping strategy to standardize the image brightness and contrast.
 
\subsection{Model Architecture and Training Configurations}\label{sec: trn-config}
To unify the experimental setup, we employed a U-Net with the EfficientNet-b2 backbone as the segmentation network, the same as for CSDG \citep{ouyang2022causality} and SLAug \citep{su2023rethinking}. Both the spatial kernel $G_{\sigma_s}$ for the spatial proximity and range kernel $G_{\sigma_r}$ for the intensity difference are set to 75 to create phase attention in Eq. \eqref{eq:bilateral}. We set $\alpha$ to 1 in the phase image reconstruction. Empirically, the standard deviations $\sigma_1$ and $\sigma_2$ in Eq. \eqref{eq:fat-bezier} were set to 0.1 and 0.5, respectively, which are the same values as used in \citet{su2023rethinking}. We set the weighting coefficient $\lambda$ for the self-Mixup strategy in Eq. \eqref{eq:fat} to 0.5. Aside from using the proposed augmentation method, we have engaged additional geometric- and intensity-based general augmentation, including affine transformation, elastic transformation, brightness, contrast, gamma transformation, and additive Gaussian noise, by default. Before augmentation, we normalized the input $\x$ using the min-max normalization with a range of [0,1]. For a fair comparison, standard data augmentations were equivalently applied to competing methods in all experiments.

We used the Adam optimizer \citep{kingma2014adam} for the model optimization and set an initial learning rate of $3\times 10^{-4}$ with a learning decay rate of $3\times 10^{-5}$. The epoch and batch size for all experiments were initialized as 2,000 and 32, respectively, and the model was assessed at the end of the training (\ie, $2,000^{th}$ epoch), where the learning decay rate converged to zero. The proposed method was implemented using the PyTorch framework and was trained on a workstation with an Intel Xeon Silver 4216 CPU @ 2.10 GHz with a single NVIDIA RTX A6000 GPU (48 GB memory).

\subsection{Evaluation Protocols}
To quantify the segmentation results, we adopted the Dice score as an evaluation metric, which is widely used in segmentation tasks. This score measures the spatial overlap accuracy between the predicted segmentation mask and ground truth, where a higher Dice score indicates better segmentation performance.

In the proposed approach to segmenting abdominal and prostate images, the dataset for the source domain was partitioned into 70$\%$ for training, 10$\%$ for validation, and 20$\%$ for testing, whereas datasets from the target domains were exclusively used for testing purposes, as in~\citet{liu2020shape}. As a special instance, the prostate dataset consists of multiple data sources (\ie, six sites), so we systematically selected one domain as the source and the remaining five domains as the evaluation targets. Therefore, the Dice score for prostate segmentation is derived from the performance average in each target domain, and such a one-versus-five experiment was conducted iteratively across all six domains. Data partitioning for cardiac cross-sequence segmentation was performed identically, following the process in \citet{su2023rethinking}.

\begin{table}[t]\centering\scriptsize\setlength{\tabcolsep}{3.5pt}
    \caption{Dice performance on abdominal and cardiac scenarios. The segmentation model for the abdominal CT-MRI scenario is trained on the CT images (\ie, the source domain) and evaluated on the MRI images (\ie, the target domain). The highest scores are in boldface, and the second-highest scores are underlined.}
    \vspace{-0.1cm}
    \label{table:abdominal_cardiac1_res}
    \centering
    \begin{tabular}{cccccccccc}
    \toprule
    \multicolumn{1}{c}{\multirow{2}{*}{\textbf{Methods}}} & \multicolumn{5}{c}{\textbf{Abdominal CT-MRI $(\%)$}} & \multicolumn{4}{c}{\textbf{Cardiac bSSFP-LGE $(\%)$}}\\
    \cmidrule(lr){2-6} \cmidrule(lr){7-10}
    & \textbf{Liver} & \textbf{R-Kid} & \textbf{L-Kid} & \textbf{Spleen} & \textbf{Average} & \textbf{LVC} & \textbf{MYO} & \textbf{RVC} & \textbf{Average} \\

\midrule
    \text { Supervised } & 91.30 & 92.43 & 89.86 & 89.83 & \cellcolor{mygray}90.85 & 92.04 & 83.11 & 89.30 & \cellcolor{mygray}88.15 \\
    \text { ERM } & 78.03 & 78.11 & 78.45 & 74.65 & \cellcolor{mygray}77.31 & 86.06 & 66.98 & 74.94 & \cellcolor{mygray}75.99 \\
    \midrule
    \text { Cutout } & 79.80 & 82.32 & 82.14 & 76.24 & \cellcolor{mygray}80.12 & 88.35 & 69.06 & 79.19 & \cellcolor{mygray}78.87 \\
    \text { RSC } & 76.40 & 75.79 & 76.60 & 67.56 & \cellcolor{mygray}74.09 & 87.06 & 69.77 & 75.69 & \cellcolor{mygray}77.51 \\
    \text { MixStyle } & 77.63 & 78.41 & 78.03 & 77.12 & \cellcolor{mygray}77.80 & 85.78 & 64.23 & 75.61 & \cellcolor{mygray}75.21 \\
    \text { AdvBias } & 78.54 & 81.70 & 80.69 & 79.73 & \cellcolor{mygray}80.17 & 88.23 & 70.29 & 80.32 & \cellcolor{mygray}79.62 \\
    \text { RandConv } & 73.63 & 79.69 & 85.89 & 83.43 & \cellcolor{mygray}80.66 & 89.88 & 75.60 & 85.70 & \cellcolor{mygray}83.73 \\
    \text { CSDG } & {86.62} & {87.48} & {86.88} & {84.27} & \cellcolor{mygray}{86.31} & {90.35} & {77.82} & {86.87} & \cellcolor{mygray}{85.01} \\
    \text { SLAug } & \textbf{90.08} & \underline{89.23} & \underline{87.54} & \textbf{87.67} & \cellcolor{mygray}\underline{88.63} & \textbf{91.53} & \underline{80.65} & \underline{87.90} & \cellcolor{mygray}\underline{86.69}\\
    \midrule
    \textbf{FIESTA (Ours)} & \underline{89.31} & \textbf{89.93} & \textbf{89.13} & \underline{87.00} & \cellcolor{mygray}\textbf{88.84} & \underline{91.42} & \textbf{81.20} & \textbf{89.22} & \cellcolor{mygray}\textbf{87.28} \\
    \bottomrule
    % \vspace{0.1cm}
    \toprule
    \multicolumn{1}{c}{\multirow{2}{*}{\textbf{Methods}}} & \multicolumn{5}{c}{\textbf{Abdominal MRI-CT $(\%)$}} & \multicolumn{4}{c}{\textbf{Cardiac LGE-bSSFP $(\%)$}}\\
    \cmidrule(lr){2-6} \cmidrule(lr){7-10}
    & \textbf{Liver} & \textbf{R-Kid} & \textbf{L-Kid} & \textbf{Spleen} & \textbf{Average} & \textbf{LVC} & \textbf{MYO} & \textbf{RVC} & \textbf{Average} \\

\midrule
    \text { Supervised } & 98.87 & 92.11 & 91.75 & 88.55 & \cellcolor{mygray}89.74 & 91.16 & 82.93 & 90.39 & \cellcolor{mygray}88.16 \\
    \text { ERM } & 87.90 & 40.44 & 65.17 & 55.90 & \cellcolor{mygray}62.35 & 90.16 & 78.59 & 87.04 & \cellcolor{mygray}85.26 \\
    \midrule
    \text { Cutout } & 86.99 & 63.66 & 73.74 & 57.60 & \cellcolor{mygray}70.50 & 90.88 & 79.14 & 87.74 & \cellcolor{mygray}85.92 \\
    \text { RSC } & {88.10} & 46.60 & 75.94 & 53.61 & \cellcolor{mygray}66.07 & 90.21 & 78.63 & 87.96 & \cellcolor{mygray}85.60 \\
    \text { MixStyle } & 86.66 & 48.26 & 65.20 & 55.68 & \cellcolor{mygray}63.95 & 91.22 & 79.64 & 88.16 & \cellcolor{mygray}86.34 \\
    \text { AdvBias } & 87.63 & 52.48 & 68.28 & 50.95 & \cellcolor{mygray}64.84 & 91.20 & 79.50 & 88.10 & \cellcolor{mygray}86.27 \\
    \text { RandConv } & 84.14 & 76.81 & 77.99 & 67.32 & \cellcolor{mygray}76.56 & \textbf{91.98} & {80.92} & 88.83 & \cellcolor{mygray}{87.24} \\
    \text { CSDG } & 85.62 & {80.02} & {80.42} & {75.56} & \cellcolor{mygray}{80.40} & 91.37 & 80.43 & {89.16} & \cellcolor{mygray}86.99 \\
    \text { SLAug } & \textbf{89.26} & \underline{80.98} & \underline{82.05} & \underline{79.93} & \cellcolor{mygray}\underline{83.05} & \underline{91.92} & \textbf{81.49} & \textbf{89.61} & \cellcolor{mygray}\textbf{87.67} \\
    \midrule
    \textbf{FIESTA (Ours)} & \underline{88.77} & \textbf{83.84} & \textbf{82.82} & \textbf{81.40} & \cellcolor{mygray}\textbf{84.21} & 91.85 & \underline{81.38} & \underline{89.56} & \cellcolor{mygray}\underline{87.60} \\
% \bottomrule
\bottomrule
\end{tabular}
\end{table}

\begin{table}[t!]\centering\scriptsize\setlength{\tabcolsep}{1pt}
    \caption{Dice performance results for the prostate dataset. The segmentation model is trained on one site (\ie, the source domain) and evaluated on the remaining sites (\ie, the Rest). The highest scores are in boldface, and the second-highest scores are underlined.}
    \vspace{-0.1cm}
    \label{table:prostate_res}
    \centering
    \scalebox{1}{\begin{tabular}{cccccccc}
    \toprule
    \multicolumn{1}{c}{\multirow{2}{*}{\textbf{Methods}}} & \multicolumn{7}{c}{\textbf{Prostate Cross-site $(\%)$}}\\
    \cmidrule(lr){2-8}
    & \textbf { A-Rest } & \textbf { B-Rest } & \textbf { C-Rest } & \textbf { D-Rest } & \textbf { E-Rest } & \textbf { F-Rest } & \textbf { Average } \\

\midrule
    \text{ Supervised } & 83.75 & 84.78 & 84.92 & 84.98 & 86.68 & 84.92 & \cellcolor{mygray}85.01 \\
    \text{ERM} & 71.81 & 65.56 & 43.98 & 71.97 & 48.39 & 37.82 & \cellcolor{mygray}56.59 \\
    \midrule
    \text{Cutout} & 78.36 & 69.08 & \textbf{63.45} & 66.39 & 61.88 & 60.19 & \cellcolor{mygray}66.56 \\
    \text{RSC} & 72.81 & \underline{70.18} & 49.18 & 74.11 & 54.73 & 43.69 & \cellcolor{mygray}60.78 \\
    \text{MixStyle} & 73.24 & 58.06 & 44.75 & 66.78 & 49.81 & 49.73 & \cellcolor{mygray}57.06 \\
    \text{AdvBias} & 78.15 & 62.24 & 54.73 & 72.65 & 53.14 & 51.00 & \cellcolor{mygray}61.98 \\
    \text{RandConv} & 77.28 & 60.77 & 53.54 & 66.21 & 52.12 & 36.52 & \cellcolor{mygray}57.74 \\
    \text{CSDG} & \underline{82.14} & 67.21 & 59.11 & 73.16 & \underline{67.38} & \underline{73.23} & \cellcolor{mygray}\underline{70.37} \\
    \text{SLAug} & 81.47 & 65.19 & 52.69 & 76.89 & \textbf{68.02} & 72.66 & \cellcolor{mygray}69.49 \\
    \midrule
    \textbf{FIESTA (Ours)} & \textbf{83.02} & \textbf{70.42} & \underline{62.06} & \textbf{77.31} & 66.17 & \textbf{74.79} & \cellcolor{mygray}\textbf{72.30}\\
\bottomrule
\end{tabular}}
\end{table}

\section{Results and Discussion}
% Shape-aware Meta-learning for Generalizing Prostate MRI Segmentation to Unseen Domains -> 실험 작성 내용 참고
\subsection{Comparison with State-of-the-art SDG Methods}
For a comprehensive comparison, we thoroughly assessed FIESTA against the empirical risk minimization (ERM) and recent state-of-the-art SDG methods, including Cutout \citep{devries2017improved}, RSC \citep{huang2020self}, MixStyle \citep{zhou2021domain}, AdvBias \citep{chen2020realistic}, RandConv \citep{xu2020robust}, CSDG \citep{ouyang2022causality}, and SLAug \citep{su2023rethinking}. Tables \ref{table:abdominal_cardiac1_res} and \ref{table:prostate_res} summarize the Dice scores for each comparative method on three cross-domain scenarios (\ie, modality, sequence, and site) in which the segmentation model is trained on the source domain data and tested on the unseen target domain data.

The numerical assessment reveals that MIS-specialized SDG methods, such as CSDG and SLAug, generally exhibited excellent performance in most scenarios. Beyond such superiority, FIESTA achieved performance on par with, or even surpassing, all baseline models, marking a substantial average improvement of 7.19$\%$ in the Dice score compared to the collective average of the baselines. Intriguingly, FIESTA underscored a significant gap in the abdominal MRI-CT and prostate cross-site scenarios, with performance boosts of 1.16$\%(\uparrow)$ and 2.81$\%(\uparrow)$, respectively, compared to the latest model, SLAug. From this perspective, exploiting FIESTA for data augmentation was especially beneficial in challenging contexts where high heterogeneity occurs, such as varying modalities or multiple sites. Moreover, RSC and MixStyle, which primarily manipulate feature representation within the latent space, demonstrated relatively inferior outcomes (\ie, limited generalizability). Based on these contradictive trends, we confirmed that leveraging data-level augmentation to improve model generalizability in cross-domain settings yields more effective results rather than it does at the feature-level representation. To compare the upper bounds further, we calculated Dice scores regarding supervised learning as a criterion, where training and testing were simulated on the target domain. Thus, the performance of FIESTA in abdominal and cardiac scenarios nearly reached the upper bounds. This finding suggests that FIESTA could alleviate the unpredictable domain discrepancy (\ie, the domain shift problem), effectively narrowing the distribution gap between the source and unseen target domains.

\begin{figure*}[t]
    \centering
    \includegraphics[width=1\linewidth]{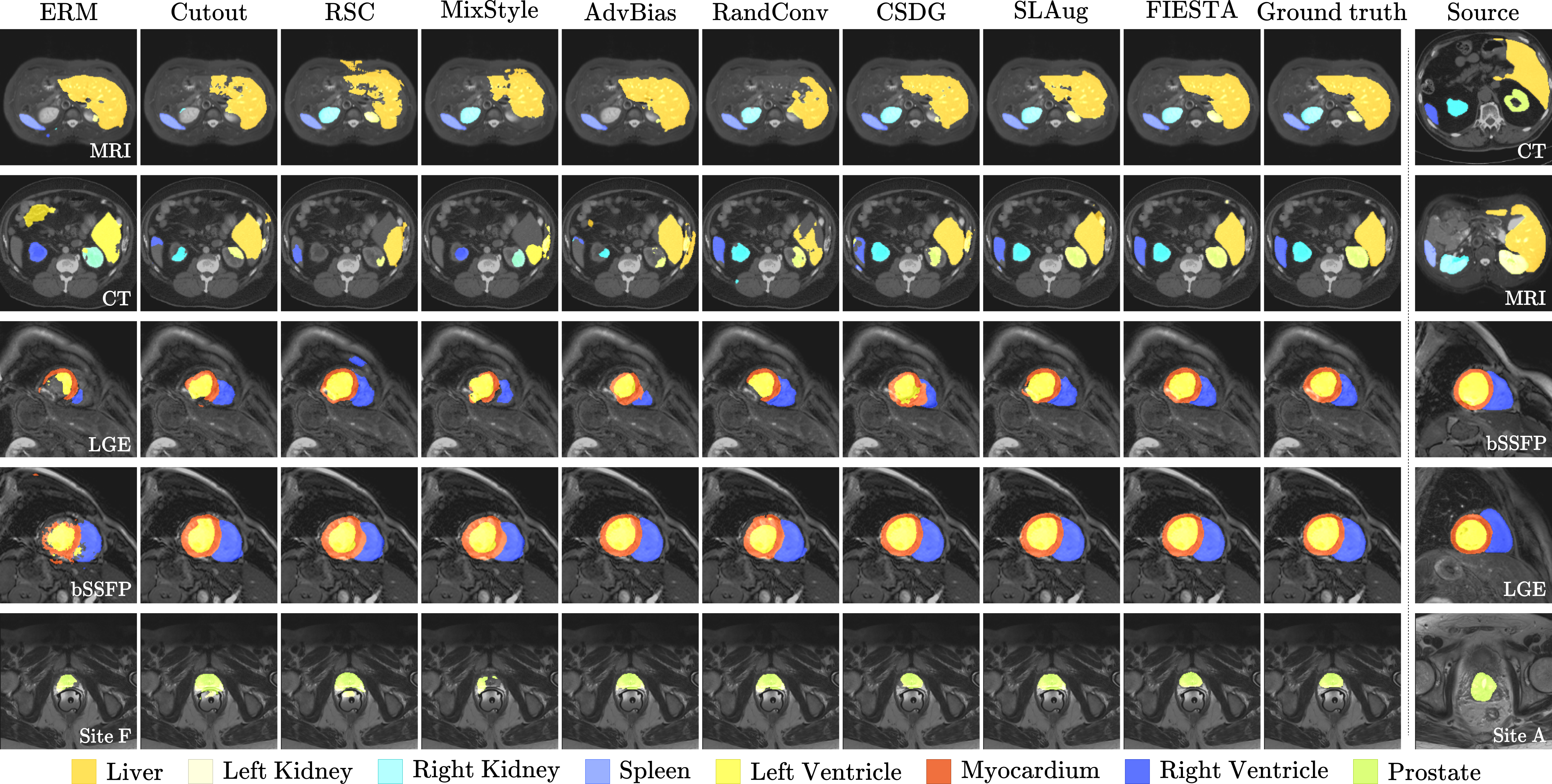}
    \caption{Qualitative analysis of three cross-domain scenarios: abdominal CT and MRI for cross-modality segmentation (Rows 1 and 2), cardiac bSSFP MRI and LGE MRI for cross-sequence segmentation (Rows 3 and 4), and prostate MRI from Site A to F for cross-site segmentation (bottom row). The rightmost images are the training data, and the remaining images represent the segmentation results of each method evaluated with the unseen target data.}
    \label{fig:sdg-qual}
\end{figure*}

Fig. \ref{fig:sdg-qual} exhibits the qualitative outcomes to evaluate the visual quality of the predicted segmentation masks. When examining the results of nonmedical approaches (\ie, Cutout, RSC, MixStyle, and RandConv), the presence of disruptive noise and the instances of mis-segmentation were confirmed, which could potentially lead to a significant performance decline. This result was particularly noticeable in the abdominal and cardiac scenarios, where the Cutout, AdvBias, and RandConv methods displayed evident problems, including partial occlusions and truncations in certain image segments. Moreover, the RSC and MixStyle methods neglected crucial segments in the abdominal images. In contrast, the medical-specific approaches of CSDG and SLAug displayed plausible portion segment regions, whereas inconsistencies were still apparent in the overall segmentation quality. Conversely, FIESTA proficiently delineated fine details and contours, producing precise segmentation outcomes closely aligned with the ground truth across varying scenarios.

\begin{figure*}[t]
    \centering
    \includegraphics[width=1\linewidth]{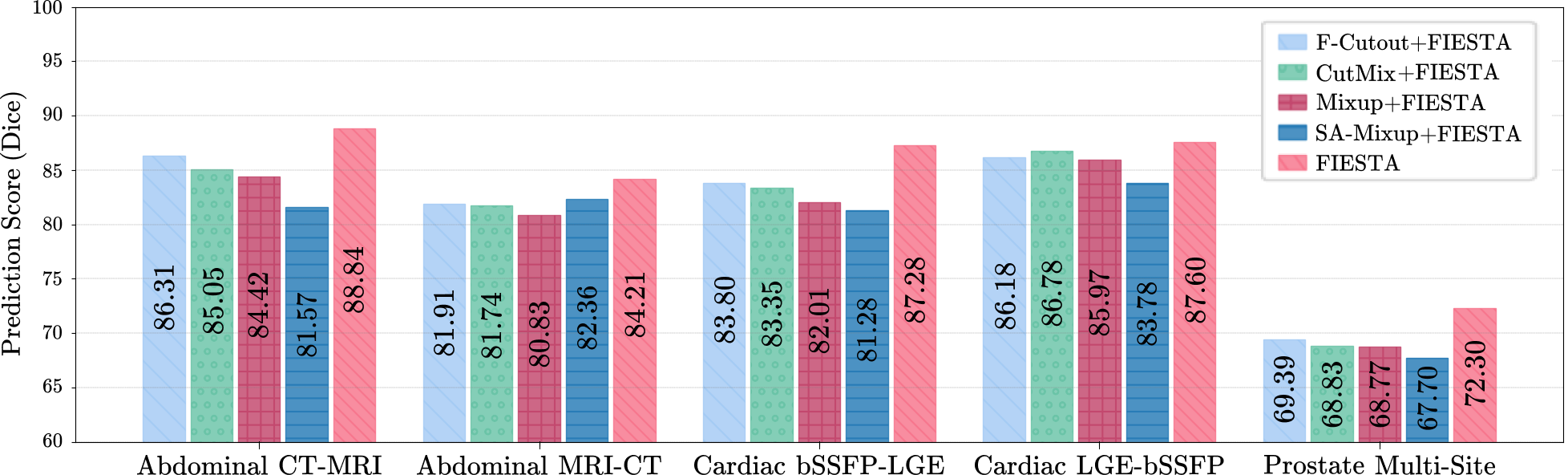}
    \caption{Comparative results of segmentation performance ($\%$) against prevalent corruption methods (\ie, F-Cutout, CutMix, Mixup, and SA-Mixup) across cross-domain scenarios. These approaches effectively manipulate the amplitude or phase component within the frequency domain.}
    \label{fig:fourier-abl}
\end{figure*}

\subsubsection{Evaluation of FAT via Fourier-Based Common Corruption}
To validate the effectiveness of the proposed FAT empirically, we conducted a series of comparative experiments against various data corruption strategies in the frequency domain. These strategies include random masking (\ie, F-Cutout\footnote{The method of F-Cutout used in this section is identical to the Cutout strategy reported in Tables \ref{table:abdominal_cardiac1_res} and \ref{table:prostate_res}, but note that it is performed in the amplitude spectrum.}), random swapping (\ie, CutMix), random blending (\ie, Mixup), and SA-Mixup—each representing conventional augmentation techniques with a stochastic component. Fig. \ref{fig:fourier-abl} presents the numerical comparison results of these approaches. To ensure a fair comparison with the FAT, we replicated the FIESTA framework, including the uncertainty-guided MA step, and replaced only the technique of manipulating the amplitude or phase component by applying these common corruptions instead of using the FAT.

In the analysis, SA-Mixup consistently yielded the lowest segmentation performance across all tested scenarios except for the abdominal MRI-CT scenario, despite being most similar to the proposed approach in terms of manipulating both amplitude and phase components. Thus, the FAT using semantic information derived by Eq. \eqref{eq:ang-den-dis} revealed a more positive contribution to model learning than SA-Mixup, which introduces a simple Mixup-based augmentation. Meanwhile, while the results between F-Cutout and CutMix generally exhibited a marginal performance gap, it was confirmed that the utilization of the F-Cutout showed a relatively predominant performance compared to CutMix in the Abdominal CT-MRI scenario. From this perspective, we assumed that random masking of certain portions within the amplitude spectrum could be a simple way to promote model robustness via missing or occluded information. 

In contrast, the Dice scores in Mixup coherently tended to yield subpar performance in all scenarios. This observation implies that the indiscriminate blending of frequency components in the amplitude spectrum might compromise its inherent characteristics, potentially generating images that stray significantly from the source data distribution. Consequently, the findings from these experiments indicate that FIESTA outperformed F-Cutout by a noticeable margin in all scenarios and improved performance beyond that achievable with generic corruption-based augmentation strategies. It allowed FIESTA to validate the feasibility of its segmentation capabilities over unseen domains. Furthermore, these outcomes underscore the faculty of FAT, which is designed to alter images systematically based on meaningful factors containing crucial information in the frequency spectrum.

\subsection{Effectiveness of Uncertainty Guidance}
Contrary to the saliency-based augmentation strategy employed in CSDG and SLAug, FIESTA strategically guides the augmentation process via epistemic uncertainty estimation. This delicate use of uncertainty differentiates FIESTA from conventional methods, optimizing the model learning by focusing on regions where the model decision is least certain.

\begin{figure}
    \centering
    \includegraphics[width=.91\linewidth]{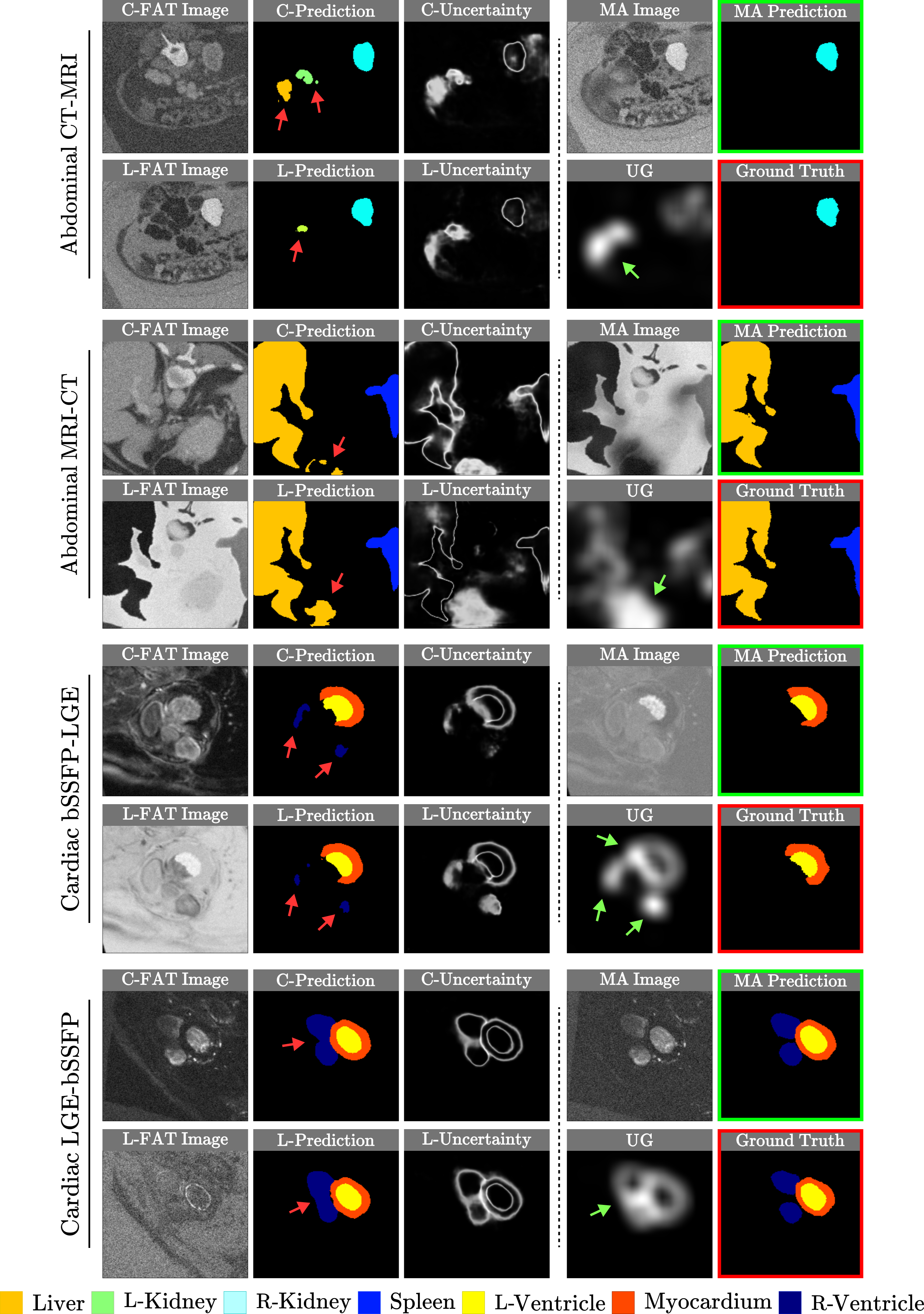}
    \caption{Visualization of uncertainty guidance (UG) and predicted segmentation output from mutually augmented (MA) images. Red and green arrows indicate mis-segmented locations and high uncertainty areas that may cause an incorrect prediction.}\label{fig:uncer-res}
\end{figure}

To ascertain the superiority of the UG, we conducted a qualitative analysis using MA images, visually examining how the segmentation quality outcomes evolve when assisted by UG (Fig. \ref{fig:uncer-res}). An interesting observation from these illustrations is the divergent outcomes of estimated uncertainty (\ie, C-uncertainty and L-uncertainty) and predicted segmentation (\ie, C-prediction and L-prediction) from the C-FAT and L-FAT images, respectively. Regions proficiently segmented in C-FAT images tend to be overlooked in L-FAT images owing to the differences in how each approach adjusts for overall or partial contrast (Fig. \ref{fig:uncer-res} Column 2). Given this, we speculated that this phenomenon may have the potential to counter domain shift challenges that inevitably degrade performance even when segmenting similar tissues or organs across modalities. 

Moreover, we confirmed that the UG could encompass an extensive array of areas subject to uncertainty contingent on contrast variations in C-FAT and L-FAT images. Ambiguous regions common to both images were relatively highlighted. Applying the UG to both types of augmented images focuses on confounding areas and introduces additional augmentation effects, establishing an enriched contrast spectrum. This process allows the segmentation model to yield outcomes more closely resembling the ground truth by properly trimming over-segmented results in cardiac LGE-bSSFP and mitigating mis-segmentation issues in abdominal scenarios and cardiac bSSFP-LGE. Consequently, we demonstrated that qualitative assessments provide clear insight into how MA images using UG could enhance the model's attention to ambiguous regions, refining its segmentation capabilities.

\subsection{Ablation Study}
A series of ablation analyses were conducted to provide insightful evidence of the efficacy of FIESTA. Table \ref{tab:abl_study} presents the structured cases for three principal components within the FIESTA pipeline: C-FAT (including amplitude masking, intra-modulation, and phase attention), L-FAT, and UG. Case 1, which does not encompass all primary components (\ie, without any checkmarks), indicates performance using only geometric and intensity-based general augmentation methods, which are described in Section \ref{sec: trn-config}.  

\begin{table}[t!]\centering\scriptsize\setlength{\tabcolsep}{3pt} % 7pt
    \caption{Ablation studies on each core component in FIESTA using abdominal CT-MRI, cardiac bSSFP-LGE, and prostate cross-site scenarios. We quantitatively evaluated each module's Dice scores ($\%$) according to the various ablation cases. Performance improvement for each case compared to Case 1 is reported in parentheses. The average performance improvement over all scenarios (\ie, Impro. Average) is revealed in the rightmost column. For brevity, we used the abbreviations for amplitude masking, intra-modulation, and phase attention as a Mask, IM, and PA, respectively.}
    \centering \label{tab:abl_study}
    \begin{tabular}{cccccccccc}
    \toprule
    \multicolumn{1}{c}{\multirow{3.5}{*}{\textbf{Case}}} & \multicolumn{5}{c}{{\textbf{Components}}} & \multirow{2.5}{*}{\textbf{Abdominal}} & \multirow{2.5}{*}{\textbf{Cardiac}} & \multirow{2.5}{*}{\textbf{Prostate}} & \multirow{2.5}{*}{\textbf{Impro.}}\\
    % \cmidrule(lr){2-6}
    & \multicolumn{3}{c}{{\textbf{C-FAT}}} & \multirow{2.5}{*}{\textbf{L-FAT}} & \multirow{2.5}{*}{\textbf{UG}} & \multirow{2.5}{*}{\textbf{MRI-CT}} & \multirow{2.5}{*}{\textbf{bSSFP-LGE}} & \multirow{2.5}{*}{\textbf{Cross-site}} & \multirow{2.5}{*}{\textbf{Average}} \\
    \cmidrule(lr){2-4}
    & \text{Mask} & \text{IM} & \text{PA} & & &\\
    \midrule
    \textbf{1} & \text{-} & \text{-} & \text{-} & \text{-} & \text{-} & 61.95 & 71.97 & 58.48 & \text{-} \\
    
    \midrule
    \textbf{2} & \multicolumn{1}{c}{\checkmark} & \text{-} & \text{-} & \text{-} & \text{-} & 67.94(+5.99) & 75.65(+3.68) & 62.43(+3.95) &\cellcolor{mygray}+4.54 \\
    
    \midrule
    \textbf{3} & \text{-} & \multicolumn{1}{c}{\checkmark} & \text{-} & \text{-} & \text{-} & 69.58(+7.63) & 76.39(+4.42) & 62.19(+3.71) &\cellcolor{mygray}+5.25 \\

    \midrule
    \textbf{4} & \text{-} & \text{-} & \multicolumn{1}{c}{\checkmark} & \text{-} & \text{-} & 66.81(+4.86) & 74.11(+2.14) & 61.85(+3.37) &\cellcolor{mygray}+3.50 \\

    % \midrule
    % \textbf{5} & \multicolumn{1}{c}{\checkmark} & \multicolumn{1}{c}{\checkmark} & \text{-} & \text{-} & \text{-} & 76.78 & 78.02 & 64.27 \\
    
    \midrule
    \textbf{5} & \multicolumn{1}{c}{\checkmark} & \multicolumn{1}{c}{\checkmark} & \multicolumn{1}{c}{\checkmark} & \text{-} & \text{-} & 78.53(+16.58) & 81.64(+9.67) & 68.55(+10.07) &\cellcolor{mygray}+12.11\\

    \midrule
    \textbf{6} & \multicolumn{1}{c}{\checkmark} & \multicolumn{1}{c}{\checkmark} & \multicolumn{1}{c}{\checkmark} & \multicolumn{1}{c}{\checkmark} & \text{-} & 82.59(+20.64) & 85.92(+13.95) & 71.24(+12.76) &\cellcolor{mygray}+15.79\\
    
    \midrule\midrule
    \textbf{Ours} & \multicolumn{1}{c}{\checkmark} & \multicolumn{1}{c}{\checkmark} & \multicolumn{1}{c}{\checkmark} & \multicolumn{1}{c}{\checkmark} & \multicolumn{1}{c}{\checkmark} & \textbf{84.21(+22.26)} & \textbf{87.28(+15.31)} & \textbf{72.03(+13.55)} &\cellcolor{mygray}\textbf{+17.04}\\
    \bottomrule
    \end{tabular}
\end{table}

Compared to the baseline in Case 1, the application of either semantic masking (Case 2) or intra-modulation (Case 3) in the C-FAT led to substantial improvements, averaging an increase of approximately 4.54$\%(\uparrow)$ and 5.25$\%(\uparrow)$ across scenarios.  This finding underscores that the strategy of amplitude manipulation in the frequency domain enhances the generalizability of the model, regardless of the cross-domain scenario. In addition, even when phase attention alone was employed (Case 4), the outcomes yielded a marginal variance from the outcomes of Cases 2 and 3. 

We assumed that highlighting the semantic appearance (\eg, the structural boundaries) during data augmentation would help enforce the segmentation ability in MIS. Incorporating masking and intra-modulation alongside phase attention (Case 5) markedly improved the average performance (abdominal: 10.42$\%$, cardiac: 6.26$\%$, and prostate: 6.39$\%$) relative to the individual components of Cases 2, 3, and 4, separately. The effectiveness of a self-Mixup strategy to obtain the C-FAT was verified, indicating that it effectively unifies each component's distinctive properties, amplifying the overall performance.

The noticeable increment in performance for Case 6, which applied C-FAT and L-FAT, suggests that offering augmented images drawn from varied perspectives could entail superb domain generalization based on the improved flexibility of the alteration in texture and contrast. Harnessing UG in Case 6 realized the best performance (\ie, FIESTA). This strategy indicates that epistemic uncertainty can be employed as a guiding principle for augmentation, effectively targeting and improving the segmentation of areas marked by ambiguity. As such, we clearly demonstrated that while each component contributes to reinforcing the quantitative figures, integrating all components adequately within FIESTA robustly boosts segmentation, maximizing performance and mitigating the domain shift.

\begin{table}[t!]\centering\scriptsize\setlength{\tabcolsep}{4.5pt}
    \caption{Ablation results by module (SAM+GT vs. SAM+Full vs. SAM+FIESTA) and model architecture (\ie, SAM vs. MedSAM) in all scenarios. The SAM+GT denotes the scenario in which the SAM is fine-tuned using ground-truth (GT) bounding boxes, representing the ``upper bound'' of the proposed method. SAM+Full is the outcome when SAM is fine-tuned using the entire image size as bounding boxes. Best outcomes (excluding the upper bound) are bold, and the improvement compared to pure SAM (\ie, Impro. over baseline) is marked with a gray-colored background.}
    \centering \label{tab:sam-finetune}
    \begin{tabular}{cccccc}
    \toprule
    \multicolumn{1}{c}{\multirow{2}{*}{\textbf{Methods}}} & \multirow{1}{*}{\textbf{Abdominal}} & \multirow{1}{*}{\textbf{Abdominal}} & \multirow{1}{*}{\textbf{Cardiac}} & \multirow{1}{*}{\textbf{Cardiac}} & \multirow{1}{*}{\textbf{Prostate}}\\
    & \multirow{1}{*}{\textbf{CT-MRI}} & \multirow{1}{*}{\textbf{MRI-CT}} & \multirow{1}{*}{\textbf{bSSFP-LGE}} & \multirow{1}{*}{\textbf{LGE-bSSFP}} & \multirow{1}{*}{\textbf{Cross-site}}\\
    \midrule
    \textbf{SAM+GT} & 92.79 & 92.80 & 90.34 & 89.93 & 92.66\\
    \cmidrule(lr){1-6}
    \textbf{SAM} & 72.16 & 73.35 & 75.61 & 74.46 & 73.58 \\
    \textbf{SAM+Full} & 78.73 & 76.14 & 81.25 & 80.46 & 77.48\\
    
    \textbf{MedSAM} & 82.85 & 80.17 & 82.57 & 82.38 & 65.39\\

    \midrule\midrule
    \textbf{SAM+FIESTA} & \textbf{84.58} & \textbf{83.54} & \textbf{86.12} & \textbf{84.09} & \textbf{81.93}\\
    \cellcolor{mygray}\textbf{Impro. over baseline} & \cellcolor{mygray}\textbf{+12.42} & \cellcolor{mygray}\textbf{+10.19} & \cellcolor{mygray}\textbf{+7.51} & \cellcolor{mygray}\textbf{+4.63} & \cellcolor{mygray}\textbf{+8.35}\\
    \bottomrule
    \end{tabular}
\end{table}

\subsection{Scalability Verification Using the Segment Anything Model}
\begin{figure}[t]
    \centering
    \includegraphics[width=1\linewidth]{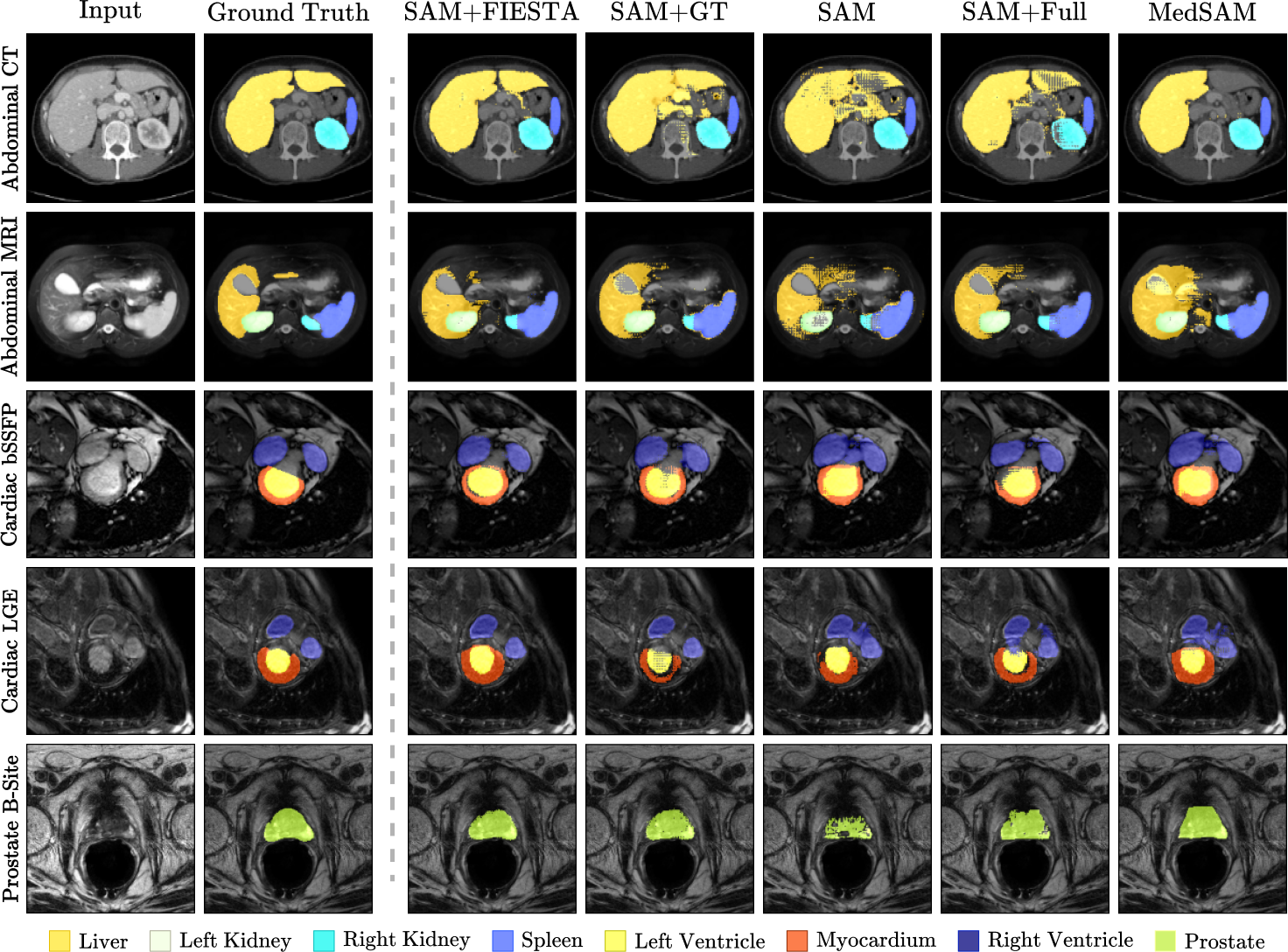}
    % \caption{Visual comparison of segmentation quality on cross-domain scenarios.}
    \caption{Qualitative visualization of the SAM-based variant methods on cross-domain scenarios.}
    \label{fig:sam_results}
\end{figure}

Last but not least, this section explores the detailed empirical analysis of the scalability of FIESTA using the segment anything model (SAM) \citep{kirillov2023segment}. The SAM is a zero-shot foundation model that achieves impressive results in image segmentation. Such favorable properties allow SAM to manage unseen datasets and tasks proficiently, such as those in SDG scenarios. However, the performance of SAM was suboptimal when delineating the intricate structure of biomedical images, where multiple organs and tissues intertwine in a single image \citep{he2023accuracy,mazurowski2023segment}. In this context, FIESTA was applied to enhance SAM further in the realm of medical images, illustrating its effectiveness and extensibility in Table \ref{tab:sam-finetune} and Fig. \ref{fig:sam_results}. Designed to be versatile, FIESTA allows seamless integration with SAM without any modifications to its architecture by virtue of its plug-and-play augmentation strategy. For a fair comparison in terms of utilizing medical images, we also employed MedSAM \citep{ma2024segment}, a variant of SAM that has been fine-tuned on immense medical datasets. 

\subsubsection{Training Protocols}
In the experimental setup, SAM and MedSAM were initialized using bounding box (Bbox) prompts as user input during MIS, adopting the vision Transformer-base model as the backbone based on each pretrained weight on the pipeline. The choice of the Bbox prompts is due to their demonstrated efficacy compared to point prompts, particularly in the context of medical segmentation \citep{huang2024segment,ma2024segment}. These underlying settings are identically applied to fine-tuning SAM via FIESTA (\ie, SAM+FIESTA). The prompt encoder was set to a fixed state because it effectively possesses the inherent capability to process the Bbox prompts. All trainable parameters in the image encoder and mask decoder were updated during fine-tuning using the FIESTA strategy.

\subsubsection{Quantitative Evaluation and Segmentation Quality Analysis}
Table \ref{tab:sam-finetune} reveals that SAM+FIESTA, involving fine-tuning SAM using the FIESTA pipeline, surpassed the performance of all other comparative methods, except for the upper bound (\ie, SAM+GT). Notably, the performance of SAM+FIESTA was close to the upper bound in cardiac scenarios, demonstrating significant enhancement. Across all tested scenarios, the average performance boost was 8.62$\%(\uparrow)$ (abdominal: 11.31$\%$, cardiac: 6.07$\%$, and prostate: 8.35$\%$) compared to using pure SAM. The SAM+Full variant yielded more improved outcomes than SAM, yet its performance was relatively inferior to MedSAM. One of the most compelling observations is that the proposed method consistently achieved superior performance against MedSAM, although MedSAM benefits from meticulously fine-tuning an extensive dataset of medical images. This superior performance suggests that fine-tuning in SAM+FIESTA, even without access to an extensive medical dataset, such as MedSAM, is a robust strategy to improve segmentation accuracy. Thus, applying an advanced fine-tuning approach with FIESTA allows similar or better outcomes for models trained on larger datasets.

As detailed in Fig. \ref{fig:sam_results}, segmentation outcomes of SAM have markedly lower quality because they consistently yield the lowest numerical scores in Table \ref{tab:sam-finetune}. The checkerboard phenomenon and mis-segmentation problem appeared in the results of the abdominal CT/MRI and prostate B-site, which could be considered a cause of performance degradation. Compared to SAM, SAM+Full alleviated the problem but still involved the occurrence of noisy and inconsistent segmentation in areas with low contrast or fragile tissue delineations. Moreover, MedSAM produced plausible visualization regarding the clarity of segment boundaries and inevitable artifacts, whereas the partial occlusion and truncation of the segment were observable in several areas, especially the liver of the abdominal CT and prostate region. 

In contrast, SAM+FIESTA provided outstanding visual quality with coherently precise segmentation output compared with the ground truth. It is worth noting that the results of SAM+FIESTA were conspicuously well-trimmed in the liver segment areas, where exaggerated segmentation or artifacts occurred in all other methods. Such quantitative and qualitative investigations highlighted the robustness and adaptability of SAM+FIESTA, suggesting that this method effectively applies the available data to achieve a significant performance leap, underscoring its potential to set new benchmarks in the MIS field.

\section{Conclusions}
This study proposes FIESTA, a Fourier-based semantic augmentation with UG, which is a novel approach designed to enhance MIS in SDG across cross-domain scenarios (\eg, abdominal cross-modality, cardiac cross-sequence, and prostate cross-site). The FAT in FIESTA strategically modulates the amplitude and phase components in the frequency domain. This manipulation allows a range of data alterations that enrich the exposure of the model to various scenarios while preserving the critical structure and semantic essence of medical images. Furthermore, we employed epistemic uncertainty estimation to guide augmentation, significantly promoting segmentation precision, especially in areas where the segmentation model struggles with ambiguous parcellated areas or indistinct tissue contrast.

Throughout rigorous assessment regimens that embrace quantitative comparison and qualitative analyses, the proposed method demonstrated significant improvement in segmentation accuracy along with visual parcellation quality compared to recent state-of-the-art SDG methods. The investigation of conventional Fourier-based corruption techniques (\eg, F-Cutout, CutMix, Mixup, and SA-Mixup) underscored the capabilities of the FAT module in manipulating the amplitude component by exploiting meaningful angular points from angular density distributions. Such distinct advantages of FIESTA in improving domain generalizability revealed that incorporating FIESTA as an advanced augmentation strategy boosts segmentation performance and the reliability of SAM, demonstrating inferior outcomes on medical images. Additionally, the proposed method adapted well to the unique textural characteristics of diverse tissues and the detailed segmentation of large and minute organs across medical imaging modalities. These findings solidify the value of FIESTA as a tool in refining MIS, demonstrating its role in domain generalization.

Moreover, FIESTA distinguishes itself through its straightforward integration (\ie, simply altering the augmentation pipeline) and impressive outcomes, facilitating considerable progress without requiring additional computational resources for modeling or more data sources. This efficiency paves the way for the inclusion of FIESTA in future SDG research endeavors in the MIS field, advocating a deeper exploration of advanced augmentation methods, such as FIESTA, to boost model robustness and accuracy in diverse medical imaging scenarios.

\section*{Code and Data Availability}
The source code for this implementation can be accessed at \url{https://github.com/ku-milab/FIESTA}. We used the abdominal cross-modality and prostate cross-site datasets, accessible at \url{https://www.synapse.org/#!Synapse:syn3193805/wiki/217789}, \url{https://chaos.grand-challenge.org/}, and \url{https://github.com/liuquande/SAML}. This repository also offers comprehensive details on the preprocessing steps and the corresponding source code. Additionally, the cardiac cross-sequence dataset was acquired from \url{https://github.com/Kaiseem/SLAug}, where it is publicly available for research purposes.

\section*{Declaration of Competing Interest}
The authors declare that they have no conflict of interest.
% The authors declare that they have no known competing financial interests or personal relationships that could have appeared to influence the work reported in this paper.

\section*{CRediT Authorship Contribution Statement}
\textbf{Kwanseok Oh:} Conceptualization, Methodology, Software, Validation, Formal analysis, Investigation, Writing - original draft preparation, Visualization, Project administration. \textbf{Eunjin Jeon} Validation, Investigation, Writing - review \& editing \textbf{Da-Woon Heo} Validation, Writing - review \& editing \textbf{Yooseung Shin} Validation, Writing - review \& editing \textbf{Heung-Il Suk:} Conceptualization, Validation, Writing - review \& editing, Supervision, Project administration, Funding acquisition.

\section*{Acknowledgements}
This work was supported by the Institute of Information \& communications Technology Planning \& Evaluation (IITP) grant funded by the Korea government (MSIT) No. 2022-0-00959 ((Part 2) Few-Shot Learning of Causal Inference in Vision and Language for Decision Making) and No. RS-2019-II190079 (Department of Artificial Intelligence (Korea University)).

\clearpage
\bibliographystyle{elsarticle-harv} 
\bibliography{main}
\end{document}